\documentclass[11pt,a4paper]{article}

\usepackage{jheppub}
\usepackage{amsmath}
\usepackage{amssymb}
\usepackage{amsthm}

\usepackage{color}

\newcommand{\thr}{\textbf{3}}
\newcommand{\thrb}{\ov{\textbf{3}}}
\newcommand{\six}{\textbf{6}}
\newcommand{\sixb}{\ov{\textbf{6}}}
\newcommand{\one}{\textbf{1}}

\newcommand{\eulc}{ u^c\, '}
\newcommand{\edlc}{ d^c\, '}
\newcommand{\eul}{ u'}
\newcommand{\edl}{ d'}
\newcommand{\etlc}{ t^c\, '}
\newcommand{\eblc}{ b^c\, '}
\newcommand{\etl}{ t'}
\newcommand{\ebl}{ b'}
\newcommand{\tQc}{ Q^c\, '}
\newcommand{\tQ}{ Q'}
\newcommand{\td}{ d'}
\newcommand{\tdc}{ d^c\, '}
\newcommand{\tu}{ u'}
\newcommand{\tuc}{ u^c\, '}
\newcommand{\tb}{ b'}
\newcommand{\tbc}{ b^c\, '}
\newcommand{\ttt}{ t'}
\newcommand{\ttc}{ t^c\, '}
\newcommand{\Ga}{G_{\text{LR}}}

\newcommand{\cM}{\mathcal{M}}

\newcommand{\cO}{\mathcal{O}}

\newcommand{\ov}{\overline}

\newcommand{\x}{\ensuremath{\times}}           
\newcommand{\ra}{\ensuremath{\rightarrow}}
\newcommand{\beq}{\begin{equation}}                                             
\newcommand{\eeq}{\end{equation}}     
\newcommand{\eeql}[1]{\label{#1}\eeq}
\newcommand{\vev}[1]{\ensuremath{\langle 0|#1|0\rangle}}
\newcommand{\veva}[1]{\ensuremath{\langle#1\rangle}}
\newcommand{\lag}{\ensuremath{\mathcal{L}}}
\newcommand{\oh}{\ensuremath{\frac{1}{2}}}                                           
\newcommand{\refl}[1]{(\ref{#1})}

\newcommand{\drawsquare}[2]{\hbox{%
\rule{#2pt}{#1pt}\hskip-#2pt
\rule{#1pt}{#2pt}\hskip-#1pt
\rule[#1pt]{#1pt}{#2pt}}\rule[#1pt]{#2pt}{#2pt}\hskip-#2pt
\rule{#2pt}{#1pt}}
\newcommand{\Ysymm}{\raisebox{-.5pt}{\drawsquare{6.5}{0.4}}\hskip-0.4pt%
\raisebox{-.5pt}{\drawsquare{6.5}{0.4}}}
\newcommand{\Yasymm}{\raisebox{-3.5pt}{\drawsquare{6.5}{0.4}}\hskip-6.9pt%
\raisebox{3pt}{\drawsquare{6.5}{0.4}}}

\newcommand{\bYasymm}{\overline{\Yasymm}}

\usepackage{color}

\title{Ultraviolet Completions of Axigluon Models and Their Phenomenological Consequences}

\author{Mirjam Cveti\v c$^{1,2}$,}
\author{James Halverson$^{1,3}$ and}
\author{Paul Langacker$^{4,5}$}

 \affiliation{$^1$Department of Physics and Astronomy, \\University of Pennsylvania,
  Philadelphia, PA 19104-6396, USA \vspace{.25cm} }
\affiliation{$^2$Center for Applied Mathematics and Theoretical Physics,\\
University of Maribor, Maribor, Slovenia \vspace{.25cm} }
 \affiliation{$^3$Kavli Institute for Theoretical Physics, \\ University of California,
  Santa Barbara, CA 93106-4030, USA \vspace{.25cm} }
 \affiliation{$^4$School of Natural Science, Institute for Advanced Study, \\ Einstein Drive, Princeton, NJ 08540, USA \vspace{.25cm} }
 \affiliation{$^5$Department of Physics, Princeton University, Princeton, NJ 08544, USA}

\emailAdd{cvetic@cvetic.hep.upenn.edu}
\emailAdd{jim@kitp.ucsb.edu}
\emailAdd{pgl@ias.edu}

\abstract{The CDF and D0 collaborations have observed a forward-backward asymmetry in $t \bar{t}$ production at large invariant mass in excess of the standard model prediction. One explanation involves a heavy color octet particle with axial vector couplings to quarks (an axigluon). We describe and contrast various aspects of axigluons obtained from the breaking of a chiral $SU(3)_L \x SU(3)_R$ gauge theory both from the standpoint of a string-inspired field theory and from a quiver analysis of a local type IIa intersecting brane construction. Special attention is paid to the additional constraints and issues that arise from these classes of top-down constructions compared with the more common effective field theory approach. These include the implications of a perturbative connection to a large scale; Yukawa couplings, which must be generated from higher-dimensional operators in many constructions; anomaly cancellation, in particular the implications of the required exotics for the axigluon width, perturbativity, and the signatures from exotic decays; the possibility of family nonuniversality via mirror representations, mixing with exotics, or additional $SU(3)$ factors; the additional constraints from anomalous $U(1)$ factors in the string constructions; tadpole cancellation, which implies new uncolored matter; the prevention of string-scale masses for vector pairs; and various phenomenological issues involving FCNC, CKM constraints, and the axigluon coupling strength. It is concluded that the construction of viable axigluon models from type IIa or similar constructions is problematic and would require considerable fine tuning, but is not entirely excluded. These considerations illustrate the importance of top-down constraints on possible TeV-scale physics, independent of the ultimate explanation of the $t \bar{t}$ asymmetry.}

\begin{document}
\begin{flushright}
{\small \tt
  \tt UPR-1244-T \\
  \tt NSF-KITP-12-171
}
\end{flushright}
\maketitle

\section{Introduction}\label{introduction}
The CDF~\cite{Aaltonen:2011kc} and D0~\cite{Abazov:2011rq}
collaborations at the Fermilab Tevatron have  reported large
forward-backward asymmetries in $t \bar{t}$ production.
For example, CDF
reports an asymmetry $A_{FB}^{t\bar{t}}= 0.296 \pm 0.067$ for
invariant mass $M_{t\bar{t}}> 450$ GeV, with an integrated luminosity
of 8.7 fb$^{-1}$. This is somewhat lower than the previous value ($0.475 \pm 0.114$ for 5.3 fb$^{-1}$), but
  is still large compared to the standard model (SM) expectation\footnote{$A_{FB}^{t\bar{t}}$ would be zero from $s$-channel gluon
  exchange in $q \bar q \ra t \bar t$ at tree level, but a small
  asymmetry is expected from higher-order effects.}: most recent calculations\footnote{However, one recent study~\cite{Brodsky:2012ik} suggests that the discrepancy may be associated with the  renormalization scale ambiguity.},  which include next-to-leading order
QCD as well as electroweak corrections,  are in the range $0.12-0.14$~\cite{Almeida:2008ug,Ahrens:2011uf,Hollik:2011ps,Kuhn:2011ri,Manohar:2012rs,Bernreuther:2012sx,Skands:2012mm,Pagani:2012kj}.

A number of possible new physics explanations have been proposed to
account for the asymmetry, including the $s$-channel exchange of a
heavy colored
particle~\cite{Ferrario:2008wm,Martynov:2009en,Ferrario:2009bz,Frampton:2009rk,Cao:2010zb,Chivukula:2010fk,Bai:2011ed,Zerwekh:2011wf,Gresham:2011pa,Djouadi:2011aj,Haisch:2011up,Tavares:2011zg,AguilarSaavedra:2011ci,Krnjaic:2011ub,Wang:2011hc},
or the $t$ or $u$-channel exchange of a $W'$, $Z'$,  Higgs, colored, or other particle~\cite{Jung:2009jz,Cheung:2009ch,Shu:2009xf,Cao:2010zb,Cao:2011ew,Shelton:2011hq,Berger:2011ua,Barger:2011ih,Bhattacherjee:2011nr,Grinstein:2011yv,Grinstein:2011dz,Patel:2011eh,Craig:2011an,Ligeti:2011vt,Gresham:2011pa,Jung:2011zv,AguilarSaavedra:2011zy,Fox:2011qd,Cui:2011xy,Duraisamy:2011pt,Blum:2011fa,Jung:2011id,Ng:2011jv,Kosnik:2011jr,Stone:2011dn,delaPuente:2011iu,Wang:2011mra,Larios:2011aa,Hochberg:2011ru,Ko:2012gj,Grinstein:2012pn,Wang:2012zv,Duffty:2012zz,Ayazi:2012bb,Hagiwara:2012gy,Allanach:2012tc,Dupuis:2012is,Ko:2012he}
with flavor changing couplings\footnote{For other mechanisms, see~\cite{Gresham:2011dg,isidori:2011dp,Foot:2011xu,Davoudiasl:2011tv,Gabrielli:2011zw,Berge:2012ih}. For effective operator
studies, see~\cite{Blum:2011up,Delaunay:2011gv,Jung:2011ym,Biswal:2012mr,Delaunay:2012kf}. For 
  reviews of the models and more complete lists of references,
  see~\cite{Cao:2010zb,Gresham:2011pa,Hewett:2011wz,Shu:2011au,AguilarSaavedra:2011ug,Gresham:2011fx,Kamenik:2011wt,Westhoff:2011tq,AguilarSaavedra:2012ma,Rodrigo:2012as}.}.  
There are stringent experimental
constraints on all of these models, especially from
the $t \bar t$ total cross section and from the LHC\footnote{LHC implications are discussed in detail in~\cite{Han:2010rf,Godbole:2010kr,AguilarSaavedra:2011vw,Hewett:2011wz,Krohn:2011tw,Haisch:2011up,AguilarSaavedra:2011hz,Gresham:2011fx,Cao:2011hr,Endo:2011tp,Knapen:2011hu,Fajfer:2012si,Ko:2012ud,Atre:2012gj,Alvarez:2012uh,bergewesthoff}.} $t\bar t$ charge
asymmetry $A_C$ (between events with the $t$ rapidity larger or smaller than that of the $\bar t$)~\cite{Chatrchyan:2011hk,ATLAS:2012an}, both of which are consistent with the SM. Other constraints include flavor changing neutral currents (FCNC), electroweak precision tests (EWPT), dijet shapes and cross sections, and $b \bar b$~\cite{Bai:2011ed,Delaunay:2012kf}
and lepton asymmetries~\cite{Bernreuther:2012sx,Berger:2011pu,Berger:2012nw,abazovD0}. The $s$-channel models are also constrained by the nonobservation of peaks in the dijet and $t\bar t$ cross sections~\cite{Chatrchyan:2011ns,Aad:2011fq,Harris:2011bh,Aad:2012wm}, while
the $t$ or $u$ channel models typically lead to observable effects in atomic parity violation~\cite{Gresham:2012wc},
associated productions such as $d G \rightarrow  t W'\rightarrow t \bar t d$~\cite{Chatrchyan:2012su},
and $tt$
production for the (Hermitian) $Z'$ models~\cite{Cao:2011ew,Berger:2011ua,Aad:2012bb}.

In particular, axigluon
models~\cite{Ferrario:2008wm,Ferrario:2009bz,Frampton:2009rk,Martynov:2009en,Chivukula:2010fk,Cao:2010zb,Bai:2011ed,Tavares:2011zg,Krnjaic:2011ub,AguilarSaavedra:2011ci,Wang:2011hc,Zerwekh:2011wf,Haisch:2011up,Djouadi:2011aj,Gresham:2011pa},
which involve an octet of heavy colored vector bosons with axial (and
possibly vector) couplings, are reasonably successful in describing
the Tevatron results, though they are strongly constrained by
other observations.  The phenomenological aspects of such models have
been extensively discussed, so here we only repeat the most essential
features. The forward-backward asymmetry is mainly due to
interference between the $s$-channel gluon and axigluon exchanges in
$q \bar q \ra t \bar t$,  which is proportional to \beq
A_{FB}^{t\bar{t}}(\hat s) \propto (\hat s - M^2) g_A^q g_A^t,
\eeql{afb} where $\hat s$ is the CM energy-squared of the $q \bar q$
system, $M$ is the axigluon mass, $g_A^q$ is the axial vector
coupling of the axigluon to light ($u$, $d$, $c$, $s$) quarks, and
$g_A^t$ is the axial coupling to the top. Most studies have assumed a
relatively heavy ($M \gtrsim (1-2)$ TeV) axigluon, so that $\hat s -
M^2 < 0$ for the relevant kinematic region. Since $A_{FB}^{t\bar{t}}
>0$ this implies opposite signs for $g_A^q$ and $g_A^t$, with
relatively large magnitudes favored, e.g., $|g_A^{q}g_A^{t}|\sim (1-4) g_s^2$, where $g_s$ is the QCD coupling. On the other hand, the nonobservation of resonant bumps in dijet production by 
ATLAS~\cite{Aad:2012wm} and CMS~\cite{Chatrchyan:2011ns} suggests (e.g.,~\cite{Bai:2011ed,Barcelo:2011vk,Choudhury:2011cg,Barcelo:2011wu}) that the axigluon should be broad,
e.g., $\Gamma/M \gtrsim 20$\%). This is to be compared to  $\Gamma/M\sim 8-10$\% expected for
decays into the ordinary quarks for axial couplings  $|g_A^{q,t,b}|=g_s$, vector couplings  $|g_V^{q,t,b}|=0$,
and $M \le 2$ TeV,
with the exact value depending on $M$ and $\alpha_s (M)$,
and suggests the need for larger couplings and/or  additional colored decay channels.
Another possibility for enhancing the asymmetry compared to the dijet constraints is to allow family nonuniversal
magnitudes, e.g., $|g_A^{q}| \ll |g_A^{t}|$~\cite{Delaunay:2011gv,Barcelo:2011vk,Delaunay:2012kf}.
Models predicting a large $A_{FB}^{t\bar{t}}$ also tend to produce a  charge asymmetry $A_C$ 
larger than observed at the LHC~\cite{AguilarSaavedra:2011hz,Chatrchyan:2011hk,ATLAS:2012an}.
The tension is currently less severe for the axigluons than for the  $t$-channel models, but nevertheless, the correlation between 
$A_{FB}^{t\bar{t}}$ and $A_C$ could be reduced for unequal couplings to $u$ and $d$ quarks~\cite{AguilarSaavedra:2012va,Drobnak:2012cz}.

One can also consider a light axigluon (e.g., $M \sim 400-450$ GeV)~\cite{Tavares:2011zg}, which allows a universal coupling
$g_A^q= g_A^t$.  In the low $M$ case, the observed asymmetry can be
obtained, and the stringent limits from the modification of the $t\bar
t$ and dijet cross sections due to axigluon exchange satisfied,
provided the coupling is relatively small, e.g., $g_A^{q,t} \sim
g_s/3$,  and the axigluon width is
large (e.g., $\Gamma/M \sim 10-20$\%). A large width and weak
coupling of course requires additional colored decay channels, and it was
assumed that the coupling is purely axial (to avoid interference
contributions to the total cross section). Very light ($M \sim 50-90$ GeV) axigluons have also been proposed~\cite{Krnjaic:2011ub}.

Most of the axigluon studies have been from the viewpoint of effective field theory, i.e., allowing arbitrary
axigluon couplings to quarks. However, additional constraints are encountered when one considers concrete theoretical embeddings.
 One possibility\footnote{Other possibilities, as well as a complete
  reference list for chiral color, can be found in~\cite{Cao:2010zb,Bai:2011ed,Atre:2012gj}.}
for obtaining an axigluon is to extend the color $SU(3)$ group to
chiral $SU(3)_L \x SU(3)_R$~\cite{Pati:1975ze,Frampton:1987dn,Frampton:1987ut,Bagger:1987fz,Cuypers:1990hb},
which is then broken to the diagonal (vector) subgroup by the vacuum
expectation value (VEV) of a scalar field transforming as
$(3,3^\ast)$.  Although such an
extension of the standard model or its minimal supersymmetric version (MSSM) is straightforward, it does introduce
some complications in terms of anomaly cancellation and for 
 the required properties for the gauge couplings and widths. In the family nonuniversal case there are additional challenges from Yukawa couplings and FCNC.
 
Further constraints arise if one attempts to obtain a chiral color extension of the MSSM from an underlying  superstring construction. 
One important consideration, which is shared in grand unification (GUT) theories or any theory involving a perturbative connection to a large scale, is the absence of Landau poles
below that scale. This strongly  restricts the possibilities for the number and type of exotic fields needed for anomaly cancellation (unless the string scale is close to the TeV scale), and also for the magnitude of the axigluon coupling. We will also assume the
absence of TeV-scale higher-dimensional operators  except for those generated by mixing with the exotic fields.

Other considerations derive from classes of superstring constructions.
We illustrate this by exploring some of the issues that arise from
embedding chiral color in a type IIa quiver\footnote{Although the
quivers we discuss have a natural home in type IIa compactifications,
these gauge theories also arise in larger classes of perturbative
string constructions via duality. This includes the T-dual type IIb
compactifications.}~\cite{Cvetic:2011iq},
which incorporates the local constraints of type IIa intersecting
brane models (see,
e.g.,~\cite{Blumenhagen:2005mu,Blumenhagen:2006ci,Cvetic:2011vz}).
Perhaps the single most important new constraint is that the field
content of the low energy theory is restricted to bifundamentals,
adjoints, symmetric, and antisymmetric representations of
non-abelian groups. Since trifundamentals are absent, tree-level quark Yukawa couplings
are often forbidden.
Furthermore, such constructions involve anomalous $U(1)$ factors that
act like effective global symmetries at the perturbative level. These
can restrict otherwise allowed couplings, and, e.g., distinguish
lepton from down-Higgs supermultiplets. There are other stringy
constraints from tadpole cancellation, the existence of a massless
(other than the Higgs mechanism) hypercharge gauge boson, and avoiding
vector pairs of fields (which typically acquire string-scale masses).
These features have phenomenological consequences.  This includes the
need to generate the quark Yukawa couplings from higher-dimensional
operators, which is especially difficult for the $t$ quark and which
leads to potential difficulties with the CKM matrix, FCNC, and
dilution of the axigluon axial coupling. The exotics required by
anomaly cancellation and other considerations can lead to a large
axigluon width, which may be desirable, but also to new constraints
from the nonobservation of their decays at the LHC. 

 In Section \ref{fieldtheory} we will review some of these issues involving chiral color from a
field theory perspective, with the additional restriction that the couplings allow a perturbative connection to a large string, compactification,  or GUT scale.
We mainly consider field representations that are easily obtainable from perturbative
string constructions, thus eliminating trifundamental Higgs fields. Quark Yukawa couplings must then be generated by higher-dimensional operators, suggesting large  ordinary-exotic mixing in the $t$ quark sector. However, for completeness we will also discuss the results of relaxing this assumption and allowing trifundamentals.
The implications for the axigluon width, effective Yukawa couplings, CKM matrix, FCNC,  exotic decays,
and the possibilitiy of generating nonuniversal couplings through ordinary-exotic mixing are described.

In Section \ref{string} we will extend the
discussion to include the additional tadpole and massless hypercharge constraints that are
required by a locally consistent type IIa intersecting brane
construction, the new colorless matter suggested by those constraints,  and  the effective global symmetries associated with the
extra anomalous $U(1)$ factors in such theories\footnote{This is part of a larger program to survey the types of extensions of the MSSM that are suggested by 
type IIa and similar constructions. See~\cite{Cvetic:2011iq} and references therein.}. 
We briefly touch on possible deformations
which could prevent string scale masses for vector pairs.
Our conclusion is that the construction of viable chiral color models from  type IIa or similar constructions
to account for the Tevatron anomaly is  problematic and would require considerable fine tuning, but is not
 entirely excluded.

Throughout,
our concern is with the general issues of constructing an axigluon
model from a consistent string construction as an illustration of the importance of top-down constraints on TeV-scale physics, independent of whether the Tevatron anomaly survives. More detailed model and
experimental issues are discussed in the references.

\section{Field Theory Construction}\label{fieldtheory}
We first consider the construction of a field theory model based on
the gauge group $\Ga\equiv SU(3)_L \x SU(3)_R\x SU(2) \x U(1)_Y$, where $SU(3)_L
\x SU(3)_R$ represents a chiral color group that will be broken to
QCD, and $SU(2) \x U(1)_Y$ is the conventional SM electroweak group.
The subscripts on the two $SU(3)$ factors indicate that the left
(right)-chiral quarks of (at least) the first two families transform
under $SU(3)_L$ ($SU(3)_R$), respectively. However, mirror families,
exotics needed for anomaly cancellation, etc., may have different
assignments.
We will label scalars and left-chiral fermion fields by $(n_L, n_R,
n_2)_Y$, where $n_L, n_R, n_2$ denote respectively the $SU(3)_L ,
SU(3)_R,$ and $ SU(2)$ representations, and $Y$ is the weak
hypercharge.  The $SU(3)_L \x SU(3)_R$ gauge interactions of quarks
are \beq -\lag= g_L \vec A^\mu_L \cdot \vec J_{L\mu} + g_R \vec
A^\mu_R \cdot \vec J_{R\mu}, \eeql{gauge} where $g_{L,R}$ are the
gauge couplings, $ A^i_{L,R}$, $i=1\cdots 8$, are the gauge bosons,
and $ J^i_{L,R}$ the currents associated with the two $SU(3)$
factors. For a single quark flavor $q$ transforming as $(3,1)+(1,3^\ast)$ the
currents are just $ J^i_{L,R\mu}=\bar q L^i \gamma_\mu P_{L,R} \, q$,
where $ L^i= \lambda^i/2$ are the $SU(3)$ matrices and $P_{L,R}=(1\mp
\gamma^5)/2$.

The simplest assignment for a single quark family that will lead to
axial vector couplings is therefore\footnote{To avoid unnecessary notation
throughout, we make the definitions $u^c \equiv u_L^c$, $d^c\equiv d_L^c$, and $e^c\equiv e_L^c$,
reserving subscripts for labeling generations. Sometimes we will use $t^c$ and $b^c$
for $u_3^c$ and $d_3^c$, depending on context.} \beq Q+ u^c + d^c=
(3,1,2)_{1/6} +(1,3^\ast,1)_{-2/3} + (1,3^\ast,1)_{1/3},
\eeql{family}
where $Q= (u \ d)^T$.
(We assume three canonical lepton families.)  The $SU(3)_L \x SU(3)_R$
symmetry can be broken to the diagonal (vector) $SU(3)$ by introducing
a Higgs field \beq \Phi=(3,3^\ast,1)_0, \eeql{Phidef} which is assumed
to acquire a VEV $\vev{\Phi}=v_\phi I$, where $I$ is the $3\x 3$
identity matrix and $v_\phi$ can be taken real and positive.  It is then straightforward to show that \beq -\lag=
g_s \vec G^\mu \cdot (\vec J_{L\mu} + \vec J_{R\mu}) + g_s \vec G_A^\mu
\cdot (\cot \delta \vec J_{L\mu} -\tan \delta \vec J_{R\mu}),
\eeql{gauge2} where $\tan \delta \equiv g_R/g_L$, $g_s\equiv g_L \sin
\delta$ is the QCD coupling, $ G^i$ are the massless QCD gluons, and $
J^i_{L\mu} + J^i_{R\mu}$ is the QCD current.  The $ G_A^i$ are the eight
axigluon fields, with mass $M_{G_A}\equiv M =\sqrt{g_L^2+g_R^2}\, v_\phi$.  In
general, the $ G_A^i$ couple to both axial and  vector currents, with the
axial ($-\gamma_\mu\gamma^5$) and vector ($\gamma_\mu$) couplings to a $(3,1)+(1,3^\ast)$ quark
given by 
\beq g_A= \oh g_s (\cot \delta+\tan \delta) \ge g_s\qquad g_V= \oh g_s (\cot \delta-\tan \delta),
\eeql{axial} 
respectively. The special case $g_L=g_R$ yields a purely axial
coupling, with $g_A=g_s$ and $M=2 g_s v_\phi$.

A family universal axigluon (for the case of a light $G_A$) can
therefore be obtained by assigning all three quark families to
transform as in \refl{family}. However, we see from \refl{axial} that
the required $g_A \sim g_s/3$ cannot be obtained at tree
level\footnote{The authors of~\cite{Tavares:2011zg} obtain $g_A< g_s$
  by introducing higher-order operators involving covariant
  derivatives, such as those that can be generated by mixing with heavy exotics~\cite{Fox:2011qd}.}.

A family nonuniversal axigluon model (for a heavy $G_A$) can be obtained
by assigning the third quark family to transform as a mirror of the
first two (e.g.,~\cite{Frampton:2009rk}), \beq t^c + b^c+
Q_3=(3^\ast,1,1)_{-2/3} + (3^\ast,1,1)_{1/3}+ (1,3,2)_{1/6},
\eeql{third} 
with $Q_3= (t\ b)^T$. This leads to $g_A^t=-g_A^q$ and $g_V^t=+g_V^q$, with $g_A^q\ge g_s$ 
and  $g_V^q$ given by
\refl{axial}.

The magnitudes of the axigluon couplings to ordinary and mirror quarks are flavor universal (e.g., $|g_A^{t}|=|g_A^u|=|g_A^d|$) at tree level in the two-node ($SU(3)_L \x SU(3)_R$) case. Flavor nonuniversality can be broken in extensions to $n$-node ($SU(3)^n$) models broken to a diagonal $SU(3)$~\cite{Chivukula:2010fk,Zerwekh:2011wf,Bai:2011ed} or by mixing effects.

There are still serious constraints from anomalies, Yukawa couplings, exotic decays,
and (for the nonuniversal case) flavor changing neutral currents
(FCNC).

\subsection{Anomaly Cancellation}\label{anomaly}
The universal and nonuniversal versions of the model both have
$SU(3)_L^3$, $SU(3)_R^3$, $SU(3)_L^2\x U(1)_Y$, and $SU(3)_R^2\x
U(1)_Y$ triangle anomalies. Various massive fermion additions to
cancel these anomalies have been suggested
(e.g.,~\cite{Frampton:1987dn}) which involve additional fermions or
families that are chiral under the electroweak part of the SM. One
common example is to introduce a fourth family which (like the third)
is a mirror of the first two with respect to $SU(3)_L \times
SU(3)_R$. However, chiral exotics would have to be very heavy to have
evaded direct detection, necessitating large Yukawa couplings to the
Higgs and therefore Landau poles at low energy
(e.g.,~\cite{Godbole:2009sy}).  While not excluded experimentally,
such Landau poles would be inconsistent with our assumption of a
perturbative extension of the MSSM up to a large string
scale\footnote{Or compactification scale, should it be small compared
to the string scale.}.  Chiral exotics also lead generically to
unacceptably large corrections to EWPT unless delicate cancellations
are present (e.g.,~\cite{Erler:2010sk,Eberhardt:2012sb}).

We therefore restrict ourselves to cancelling the anomalies using quasichiral exotic fermions.
These are defined as fermions that are vector under the SM gauge group, though they may be  chiral under additional symmetries such as
 $SU(3)_L \times SU(3)_R$. 
Such quasichiral fermions  do not require large Yukawa couplings and do not contribute at one loop to EWPT\footnote{Exotic $SU(2)$ multiplets would contribute to the $\rho$ parameter if they are nondegenerate~\cite{Beringer:1900zz}.}. They will typically acquire masses near the
chiral (in this case  $SU(3)_L \times SU(3)_R$) breaking scale.
For simplicity\footnote{Larger representations would also lead to the divergence of $g_{L}$ and/or $g_R$  at low scales.} (and to allow generation of exotic masses within our assumptions) we will only consider quasichiral exotics
that are fundamental or antifundamental under the $SU(3)_{L,R}$ factors (as well as the bifundamental Higgs fields $\Phi$ and $\ov \Phi \equiv (3^\ast,3,1)_0$).
There are two very simple  possibilities, which as far as we are aware have not been discussed in the literature:
the anomalies for an ordinary family can be cancelled by adding 
a pair of exotic
$SU(2)$ doublets $Q^c\,' =(\edlc\ -\eulc)^T$ and  $Q' =(\eul \ \edl)^T$ transforming as 
\beq Q^c\,' +Q' =
(3^\ast,1,2)_{-1/6} +(1,3,2)_{1/6}, \eeql{affamily}
or by adding two pairs of $SU(2)$ singlets
\beq \eulc+\edlc+\eul+\edl =
(3^\ast,1,1)_{-2/3}+(3^\ast,1,1)_{1/3}+(1,3,1)_{2/3}+(1,3,1)_{-1/3}.
\eeql{singletexotics}
The anomalies for a  mirror family could be cancelled by 
 \beq Q_3' + Q_3^c\,'=
(3,1,2)_{1/6}+(1,3^\ast,2)_{-1/6} \eeql{affamily2}
or 
\beq \etl+\ebl+\etlc+\eblc= 
(3,1,1)_{2/3}+(3,1,1)_{-1/3}+(1,3^\ast,1)_{-2/3}+(1,3^\ast,1)_{1/3}.
\eeql{singletexotics2}
Of course, one can also have hybrid situations, involving some singlet  and some doublet pairs. 
In the nonuniversal case, the anomalies of the third family cancel those of one of the light families,
so it would suffice to only add one family of exotics, either \refl{affamily} or \refl{singletexotics}.

\begin{figure}[htbp]
\begin{center}
\includegraphics*[scale=0.6]{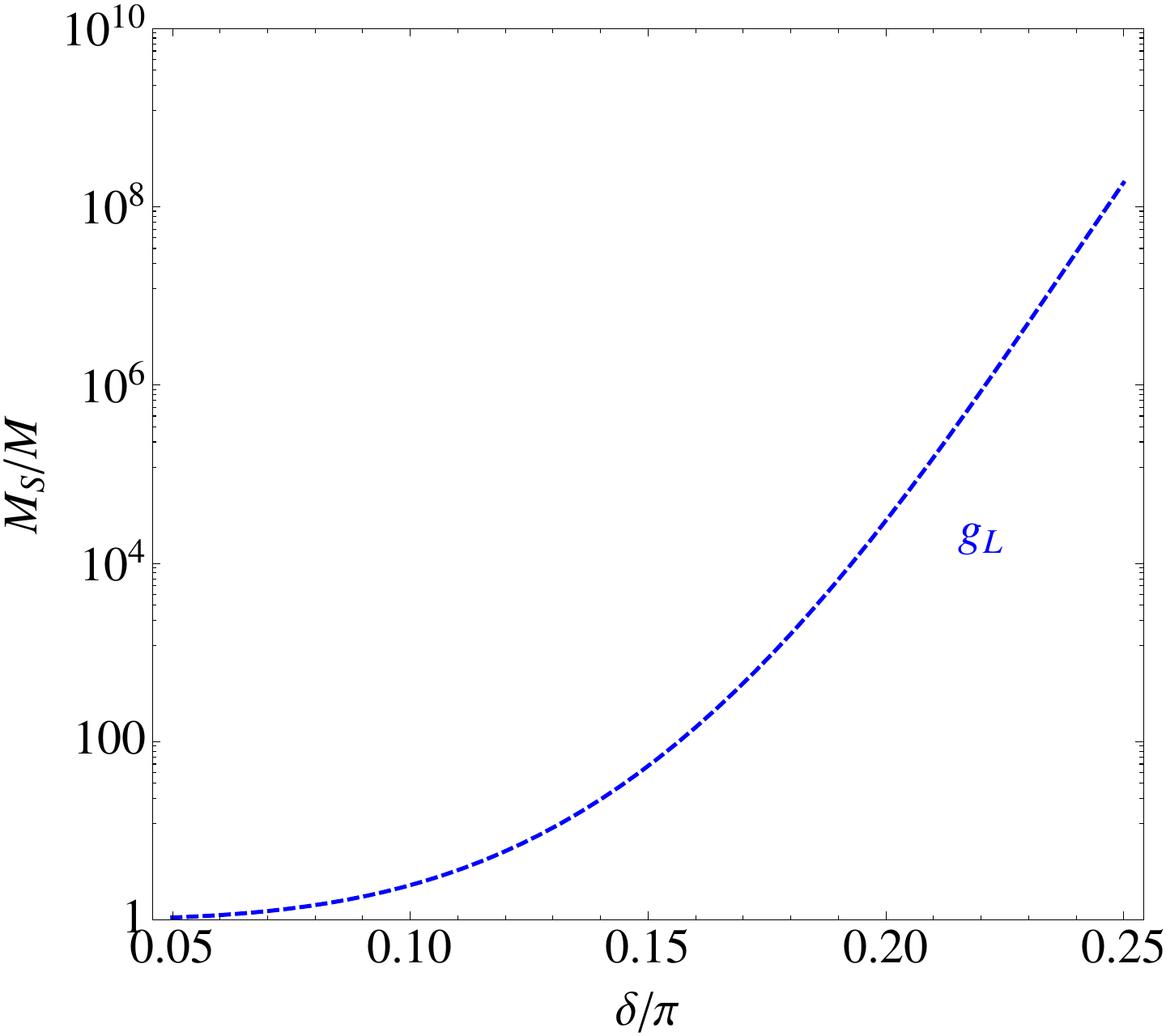}
\caption{The maximum value of  $M_S/M$ allowed by the assumption of perturbativity
of $g_L$  up to the string scale $M_S$ as a function of $\delta=\tan^{-1} (g_R/g_L)$,
assuming three families of exotics plus  $\Phi$ and $\ov \Phi$.
The mixed two-loop $\beta$ function terms of $\mathcal{O}(g_L^5)$ and $\mathcal{O}(g_L^3 g_R^2)$
 are included, but the electroweak and Yukawa couplings are neglected.
For $M=\mathcal{O}\,$(TeV) the maximum allowed string scale  is only around $10^{11}$ GeV,
and requires $\delta \sim \pi/4$ so that $g_L \sim g_R \sim \sqrt{2} g_s$ at $M$. 
The curve is symmetric around $\delta=\pi/4$, except that the roles of $g_L$ and $g_R$ are interchanged for $\delta > \pi/4$.}
\label{rge}
\end{center}
\end{figure}

There are some advantages to assuming  quasichiral exotics  for each family, i.e., they  may allow a larger axigluon width (if the exotics are sufficiently light), and they also facilitate the generation of effective Yukawa couplings in the present framework.
However, they are problematic for our assumption of perturbativity of the gauge couplings
$g_L=g_s/\sin \delta$ and $g_R=g_s/\cos \delta$. Assuming  three families of  exotics, as well as $\Phi$ and $\ov \Phi$, the  supersymmetric  $\beta$ functions for $SU(3)_L$ and $SU(3)_R$
vanish at one-loop (see Appendix \ref{rgeappendix}). However, they are positive at two loops~\cite{Jones:1981we,Einhorn:1981sx}, so that perturbativity up to the
string scale $M_S$ requires a relatively low string scale and small values of $g_{L,R}$ at $M$.
For $\delta=\pi/4$, one requires\footnote{Such intermediate scales are known to exist in
stabilized type IIb compactifications \cite{Balasubramanian:2005zx}.} $M_S \lesssim 2\times  10^{8} M$, while smaller or larger $\delta$
(useful for phenomenological reasons as described below) require a much lower string scale\footnote{The more complicated quasichiral exotics in a supersymmetric version of the two-node model in~\cite{Tavares:2011zg} would exhibit Landau poles at a still lower scale, $\sim 20 M$.}, as shown in 
Figure \ref{rge}. We will mainly consider $0.2 \lesssim \delta/\pi \lesssim 0.3$, so that $M_S/M \gtrsim 10^4$.

On the other hand, for a single family of exotics in the  nonuniversal case, as well as  $\Phi$ and $\ov \Phi$,
the $SU(3)_{L,R}$ couplings are asymptotically free, allowing $M_S$ as large as the Planck scale, provided that the initial values at $M$ are not too large.
The latter restriction is satisfied for $0.1 \lesssim \delta/\pi \lesssim 0.4$.
 We will discuss both the three  and one exotic family cases below.

As will be discussed in Section \ref{effectiveyukawas}, the $SU(2)$-doublet exotics are problematic for the third family,
because they  lead to unacceptable modifications of  the CKM matrix. 
All of the possibilities involve some pairs that are totally vector  under
$\Ga$,
such as intrafamily terms $Q + Q^c\,'$ or  $Q_3^c\,' + Q_3$, cross family terms like  $Q'+Q_3^c\,'$ or $Q_3'+Q^c\,'$,   or analogous pairs involving
$SU(2)$-singlet exotics.  We must assume that
some additional symmetries prevent such vector pairs from combining
to form very heavy (e.g., string-scale) states.
The role of the extra stringy conditions for these
requirements will be discussed in Section \ref{string}.

\subsection{Exotic Masses and Effective Yukawa Couplings}\label{effectiveyukawas}

\subsubsection*{Exotic Masses}
Another important complication involves the 
 generation of masses for the
ordinary and exotic quarks.  
Clearly, the exotic doublet pair $Q'+ Q^c\,' $ can
acquire a mass at the $SU(3)_L \x SU(3)_R\, $-breaking scale $v_\phi$ by
a renormalizable-level Yukawa coupling $ \lambda_{Q'}\Phi\,  Q' Q^c\,' $, 
while exotic singlets can acquire masses via  $\lambda_{u'} \Phi\, u' u^c\,' $
or  $\lambda_{d'}\Phi\,  d' d^c\,' $. Similarly, mirror exotics can acquire masses
from couplings to an additional field\footnote{We assume $\vev{\ov \Phi}=v_{\bar \phi} I$,
so that $M^2 =(g_L^2+g_R^2)\, (|v_\phi|^2+ |v_{\bar \phi}|^2)$. Note, however,  that $\Phi$ and $\ov \Phi$ form
a vector pair under $\Ga$, as will be discussed in Section \ref{string}.
} $\ov \Phi$ transforming as $(3^\ast,3,1)_0$, i.e.,
$\lambda_{Q_3'} \ov \Phi\,  Q_3' Q_3^c\,'$, $\lambda_{t'}  \ov \Phi\,  t' t^c\,' $,
or  $\lambda_{b'} \ov \Phi\,  b' b^c\,'$. In a nonsupersymmetric theory, the role of $\bar\Phi$ can be played by $ \Phi^\dagger$.

\subsubsection*{Trifundamental Higgs Doublets}
However, mass terms for the ordinary quarks are  problematic.
 Renormalizable-level Yukawa
couplings  $ H_u Q \,u^c$ or $ H_d Q\, d^c $ would require  Higgs
doublets transforming as the trifundamental representation $(3^\ast, 3, 2)_{\pm 1/2}$.
As emphasized in the Introduction
the trifundamental Higgs representations cannot be
obtained in the types of string constructions we are considering.
The assumption is generically valid in perturbative type II and type I 
string theory, and also in some heterotic compactifications. In F-theory
it is possible to obtain trifundamental representations from three-pronged
string junctions, but obtaining these representations as massless degrees of
freedom is relatively difficult, since they exist at high codimension in moduli space.

However, for completeness, we briefly comment on the implications of trifundamental Higgs doublets. 
The main drawback is that such states would require a very complicated construction, even at the field theory level.
A second set of Higgs doublets  $(1,1,2)_{\pm 1/2}$ \emph{not} carrying  $SU(3)_L \x SU(3)_R$ charges 
 would still be required for the Yukawa couplings of leptons, while a third set   $(3, 3^\ast, 2)_{\pm 1/2}$
 would be required for the Yukawa couplings of a mirror generation.
 It would be nontrivial to arrange for all of these fields to
 have nonzero VEVs, though it might be possible if the different sets of doublets were connected by cubic interactions involving $\Phi$ and $\ov \Phi$. On the other hand, trifundamental Higgs fields would allow straightforward generation of the ordinary quark masses, with little mixing with exotic quarks  required. Since such mixings tend to dilute the axial couplings, such constructions would have a better chance of describing the collider data.
 
\subsubsection*{$(1,1,2)_{\pm 1/2}$ Higgs Doublets}
Now consider the more challenging situation in which there are no trifundamental Higgs fields. It is still possible to generate effective
Yukawa interactions from higher-dimensional operators obtained by  integrating out
the heavy exotics (similar situations have been considered by many authors, e.g.,~\cite{Fox:2011qd,Tavares:2011zg,Shelton:2011hq}).
However, there are important complications and constraints. First, it is  nontrivial to obtain a large enough top quark mass.
Secondly, renormalizable Yukawa interactions of $(1,1,2)_{\pm 1/2}$ Higgs doublets connecting ordinary and mirror families are allowed by the gauge symmetry. 
Unless they  are suppressed by additional quantum numbers or  other means
they could lead to large CKM mixings for the third family.
Finally, bare mass terms  connecting normal   and exotic quarks are also allowed by 
$\Ga$. Unless these are forbidden or suppressed they
 can lead to  mixing effects that can considerably reduce the axial vector couplings of the  mass-eigenstate quarks to the $G_A$. Large mixings of the top quark with exotics are hard to avoid  within the present framework in the absence of trifundamentals, while they are model dependent for the other quarks. In addition to suppressing the axial couplings, mixing of the quark doublets $Q_i$ with exotic states leads to violation of CKM universality or observations, so any left-handed quark mixings must be very small. Mixing of the $Q_i$ with singlet exotics would also lead to off-diagonal couplings of the $Z$ and Higgs. Mixing of the antiquarks (i.e., of the right-handed quarks) with exotics is not constrained by observations of the CKM matrix, but  for doublet exotics would lead to off-diagonal $Z$ and Higgs couplings and right-handed couplings of the $W$.

\subsubsection*{Three Exotic Families}
We first discuss the case of three families of exotics.
Consider, for example, the $u$ quark in the model with singlet exotics, as in \refl{singletexotics}. One can write renormalizable mass and mixing terms
\beq
-\lag=\mathcal{M}_s \eul \eulc  + h_s H_u Q \eulc  + \delta_s \eul u^c, 
\eeql{umass}
where $\mathcal{M}_s \equiv \lambda_{u'} v_\phi$, 
$h_s$ is a Yukawa   coupling which connects $Q$ to $\eulc$, and $H_u$ is a Higgs doublet transforming as $(1,1,2)_{1/2}$
with VEV $v_u \equiv  \veva{H^0_u}\sim 174$ GeV $\sin \beta$, where $\tan \beta \equiv v_u/v_d$. The subscripts indicate the presence of singlet exotics, and flavor indices are suppressed.
The last term  is an $SU(3)_L \x SU(3)_R\x SU(2) \x U(1)_Y$-invariant  mixing. The coefficient $\delta_s$ could in principle be a bare mass. However, as mentioned above, it is more plausible that $\delta_s$ is generated by the VEV of a singlet field that breaks
some additional symmetry. We will assume that $\delta_s$ is smaller than or of the same order as the chiral $SU(3)$ breaking.
For $\mathcal{M}_s$ much larger than $ \delta_s$ and $h_s v_u$ one can integrate out the exotic fields  to obtain the effective Yukawa interaction $(h_s \delta_s/\mathcal{M}_s)  H_u Q u^c$, so that the $u$-quark mass is
$m_u \sim (h_s \delta_s/\mathcal{M}_s) v_u$. From the $2 \x 2$ mass matrix, the left mixing $\theta_L$ (between $u$ and $\eul$)
and right mixing $\theta_R$ (between $u^c$ and $\eulc$) angles are given by
\beq \theta_L \sim \frac{h_s v_u}{\mathcal{M}_s} \sim \frac{m_u}{\delta_s}, \qquad \theta_R \sim \frac{\delta_s}{\mathcal{M}_s}. \eeql{umixing}
In principle, $\theta_L$ and the analogous mixings for $d$, $s$, and $b$ lead to apparent violation of the
CKM universality condition
\beq  |U_{ud}|^2+ |U_{us}|^2+|U_{ub}|^2=1. \eeql{ckmuniversality}
However, the uncertainty in the observed value $\sum_i |U_{u d_i}|^2=0.9999(6)$~\cite{Beringer:1900zz}
allows  for $\theta_L$  as large as 0.02  (e.g.,~\cite{Langacker:1988ur}), with a similar limit for the $d$ quark and even weaker
limits for $s$ and $b$. For $m_u\sim 5$ MeV this is satisfied for the very weak restriction $\delta_s\gtrsim 250$ MeV. The analogous constraints for the $d$, $s$ and $b$ mixing parameters are respectively 250 MeV, 1 GeV, and 5 GeV. Since $\theta_L$ describes mixing between a doublet and singlet it induces FCNC between the light and heavy states. However, FCNC between light states
such as $u$ and $c$ is at most of second order, but can actually be  negligibly small~\cite{Langacker:1988ur}. $\theta_L$ also allows for exotic decays
into light quarks (Section \refl{width}). $\theta_R$ has no effect on SM physics, but does slightly modify the axigluon couplings.

For doublet exotics, \refl{umass} is replaced by
\beq
-\lag=\mathcal{M}_d \tQ \tQc  + h_d H_u \tQ u^c  + \delta_d Q \tQc, 
\eeql{Qmass}
where $\mathcal{M}_d=\lambda_{Q'} v_\phi$ and the subscripts refer to doublet. The lightest
mass eigenvalue is $m_u\sim h_d v_u \delta_d / \cM_d$. This is very much like the singlet case, except that the left and right mixings are interchanged, i.e., $\theta_L\sim \delta_d/\mathcal{M}_d$, $\theta_R \sim m_u/\delta_d$, and any FCNC is in the right ($u^c-\eulc$) sector. The CKM universality constraint on $\theta_L$ is satisfied for $ \delta_d/\mathcal{M}_d\lesssim 0.02$  for $u$ and $d$, and $\lesssim 0.1$ for $s$.

It is therefore straightforward to generate masses for the first two quark families, as well as the $b$ quark and Cabibbo mixing, by higher-dimensional operators. The top quark, however, is  more challenging.
Suppose, for example, that the third family is a mirror with singlet exotics, as in \refl{third} and \refl{singletexotics2}.
The relevant mass terms are
\beq
-\lag=\mathcal{M}_t \etl \etlc + h_t H_u Q_3 \etlc  + \delta_t \etl t^c, 
\eeql{tmass}
where $\mathcal{M}_t= \lambda_{t'} v_{\bar \phi}$. Assuming that $\mathcal{M}_t \gg h_t v_u$  the smaller mass eigenvalue is
$m_t \sim h_t v_u \delta_t/\sqrt{\delta_t^2+ \mathcal{M}_t^2}$. 
Unless $h_t v_u$ is quite large this requires a large value for $\delta_t $ and leads to significant mixing.
The right-handed component of the lightest mass eigenstate is 
\beq t^c_1= \cos \theta_R t^c-\sin \theta_R \etlc \sim \frac{\mathcal{M}_t}{\sqrt{\delta_t^2+ \mathcal{M}_t^2}} t^c- \frac{\delta_t}{\sqrt{\delta_t^2+ \mathcal{M}_t^2}}\etlc.
\eeql{tmixing}

In principle $h_t $ is bounded by the requirement  that there are no Landau poles
up to a large scale. However, as discussed  in Appendix  \ref{rgeappendix}  this is difficult to quantify in this case because
of the vanishing of the $\beta$ functions for $g_{L,R}$ at one loop and the rather low string scales shown in Figure  \ref{rge}.
 For the illustrative range 170 GeV $\lesssim h_t v_u \lesssim$ 450 GeV, with $m_t\sim 164$ GeV (the $\overline{MS}$ mass), $\cos \theta_R$ increases from $\sim 0.3$ to $\sim  0.9$.  Note that $t_1^c$ is   dominated by $\etlc$ rather than $t^c$ for much of the 
 range (up to $h_t v_u \sim 280$ GeV).
 The left-handed mixing is typically small\footnote{The situation would be reversed for third family doublet exotics, leading to  $V_{tb}\sim \cos \theta_L$. This contradicts the experimental result $|V_{tb}|> 0.93$~\cite{Beringer:1900zz}
unless $ \cos \theta_L$ is large enough, which would require $h_d v_u \gtrsim 450$ GeV for the third family analog of \refl{Qmass}.}.
The axial vector coupling of the axigluon to the top quark is diluted compared to \refl{axial} by the mixing. Ignoring the left-handed mixing, the reduction factor is $\cos^2 \theta_R$ (see Appendix \ref{mixingcouplings}), reducing   $A_{FB}^{t\bar{t}}$ in \refl{afb} from interference by the same amount.

We reiterate that for a mirror third family mixing terms such as $H_u Q t^c,$ $ H_u Q_3 u^c,$ $ H_d Q b^c,$ and $H_d Q_3 d^c$
are allowed by $SU(3)_L \x SU(3)_R\x SU(2) \x U(1)_Y$. This is unnatural in that it is backward from observations, i.e., the CKM mixings between the third family and the first two are tiny, suggesting small mixing terms compared to the masses. Such terms  would have to be strongly suppressed by new symmetries or other mechanisms such as world-sheet instanton effects to avoid large  third family mixing and also the dilution of the axial axigluon couplings. Similarly, if an ordinary and mirror exotic family
were both singlets or both doublets they
would involve vector pairs such as $u'+t^c\,'$  that could acquire large masses unless they are suppressed.

\subsubsection*{One Exotic Family}
Now consider the possibility of a single exotic family in the nonuniversal case\footnote{This model is actually obtained from the nonuniversal model with three singlet exotic families  if one allows the vector pairs of exotics to obtain a large mass.}. We only consider singlet exotics for the same reason as
in the three family case, i.e., because doublet exotics would lead to unacceptably  large left-handed top mixing. 
We therefore have two ordinary families $ Q_i+ u_i^c + d_i^c,\, i=1,2$ of the type in \refl{family}, one mirror family
$ Q_3+ u_3^c + d_3^c$, as in \refl{third}, and one exotic family $\eulc+\edlc+\eul+\edl$ as in \refl{singletexotics}.
Note that $ u_3^c$ and  $\eulc$ have the same quantum numbers, as do $ d_3^c$ and  $\edlc$. We use the label $ u_3^c$
rather than $t^c$ because the mass eigenstate antitop will actually consist mainly of $ u_2^c$ and  $\eulc$.

The allowed mass terms in this case are
\beq
-\lag=\mathcal{M} \eul \eulc  +  h_i H_u Q_i u_3^c + k_i  H_u Q_i  \eulc + 
\kappa_i H_u Q_3 u_i^c +  \delta_i \eul  u^c_i,
\eeql{onef}
where $\mathcal{M} \equiv \lambda v_\phi$; the $\delta_i$ are bare masses or generated by singlet VEVs; and
$h_i$, $k_i$, and $\kappa_i$, $i=1,2$, are Yukawa couplings which we assume to be perturbatively bounded.
We have used the equivalence of  $ u_3^c$ and  $\eulc$ to eliminate a coupling $\mathcal{M}' \eul u_3^c$.
Consider first the limit in which $\mathcal{M},\, \delta_2 \gg \kappa_2 v_u$ and the other couplings are neglected. This is effectively the $t$-exotic system, with the lighter of the two nonzero masses 
$m_t \sim \kappa_2 v_u \mathcal{M}/\sqrt{\delta_2^2+ \mathcal{M}}$. 
The lighter  eigenstate is $t_{1} \sim t  \in Q_3$ with little left-mixing, and 
\beq t^c_1= \cos \theta_R  \eulc -\sin \theta_R\ u_2^c \sim \frac{\delta_2}{\sqrt{\delta_2^2+ \mathcal{M}^2}} \eulc- \frac{\mathcal{M}}{\sqrt{\delta_2^2+ \mathcal{M}^2}} u_2^c.
\eeql{tonemixing}
This closely resembles \refl{tmixing}  except that $\delta$ and $\mathcal{M}$ are interchanged.
The axial coupling of the $t_1$ is $- g_A \cos^2 \theta_R$.
As discussed in Appendix \ref{rgeappendix} the absence of a Landau pole up to $M_S/M \sim 10^{15}$ leads to an upper bound on $\kappa_2 v_u$ which increases from $\sim 170$ to  $\sim 450$ GeV as $\delta/\pi$ increases from $0.1$ to $0.4$. Similar to
the three exotic family case, this implies an upper bound on $\cos \theta_R$ which increases roughly linearly 
from 0.3 to 0.9 for $0.1 < \delta/\pi < 0.4.$

We have verified numerically that the other parameters in \refl{onef} can be adjusted to give reasonable values
for the $u$ and $c$ quark masses, and that similar considerations can yield appropriate masses for the down-type quarks.
The observed  CKM mixing between the left-handed quarks can be reproduced (with the mixings occurring in either the up or down sector, or in a combination), with small left-mixing with $\eul$. Finally, the parameters can be chosen to maximize
the axial couplings of the $u$ quark (so that its right-component is $\sim u^c_1$), while the c-quark can have nearly vector couplings (i.e., its right-component can be $\sim u^c_3$).  Similar statements apply in the down sector.

\subsection{Flavor Changing Neutral Currents} \label{fcnc}

FCNC are a serious concern for the nonuniversal model, because any
mixing between the third and light fermion families would break the
GIM mechanism and generate couplings like $\bar s \gamma_\mu d$, $\bar b \gamma_\mu d$, $\bar b \gamma_\mu s$,   or $ \gamma_\mu \rightarrow  \gamma_\mu  \gamma^5$ to the axigluon, leading to new
contributions to neutral $K$, $B_d$ and $B_s$ meson mixing.  Assuming that
the observed CKM mixing is mainly due to the $d, s,$ and $b$ sector,
the observed $K$ and $B$ mixing eliminates most but not all of relevant the
parameter space~\cite{Chivukula:2010fk,Bai:2011ed,Haisch:2011up}. 
  This would be relaxed if the CKM mixing is dominantly
due to mixing between the $u, c,$ and $t$~\cite{Bai:2011ed,Haisch:2011up}
because of the weaker constraints on $D$ mixing.

Normal-exotic quark mixing can also induce FCNC between light quarks at second order, but as commented in Section \ref{effectiveyukawas} this can be negligibly small. We have not investigated  FCNC
 mediated by the
various Higgs fields  in detail.

\subsection{The Axigluon Width, Exotic Decays, and Universality}\label{width}

\begin{figure}[htbp]
\begin{center}
\includegraphics*[scale=0.5]{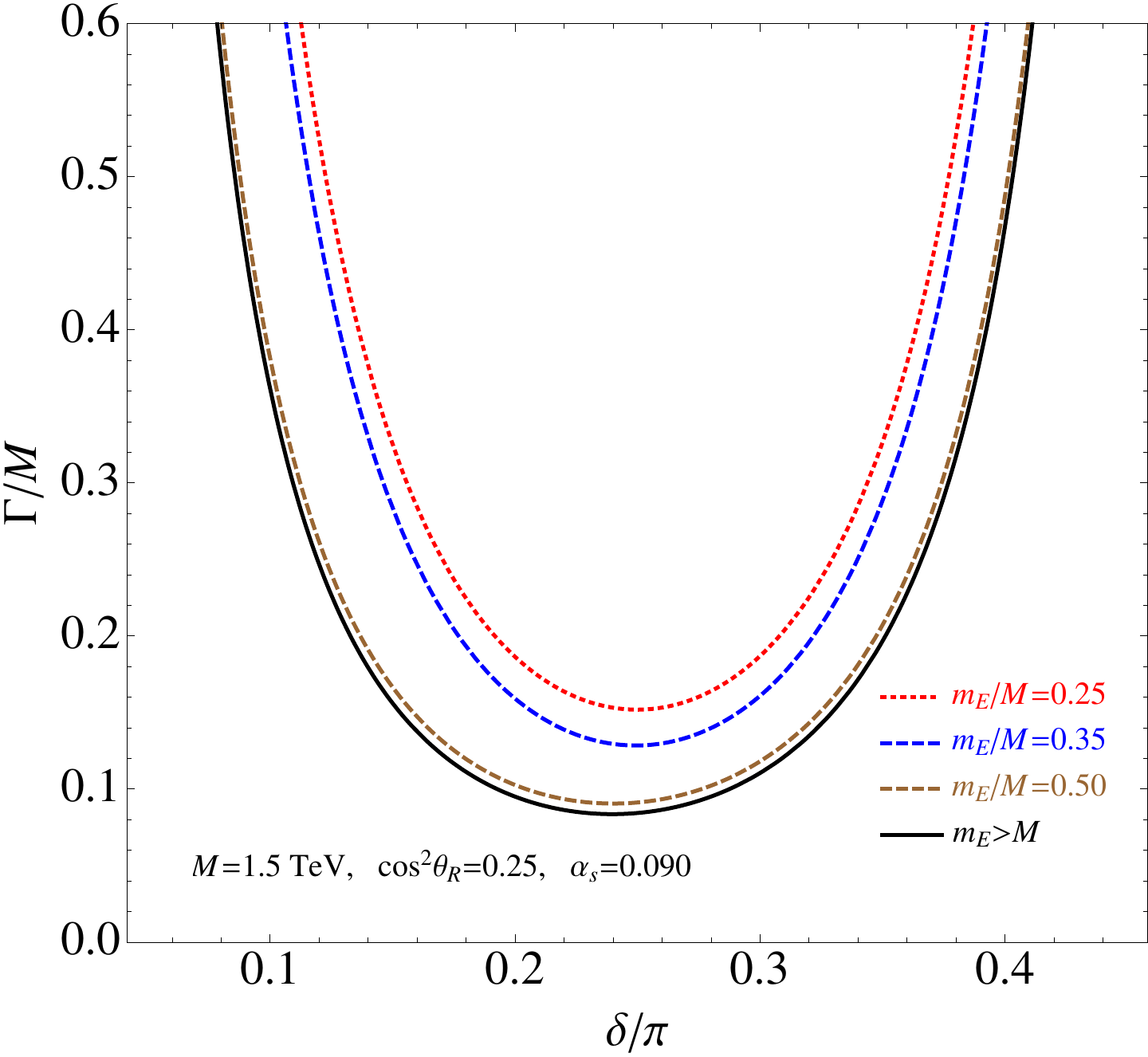}
\includegraphics*[scale=0.5]{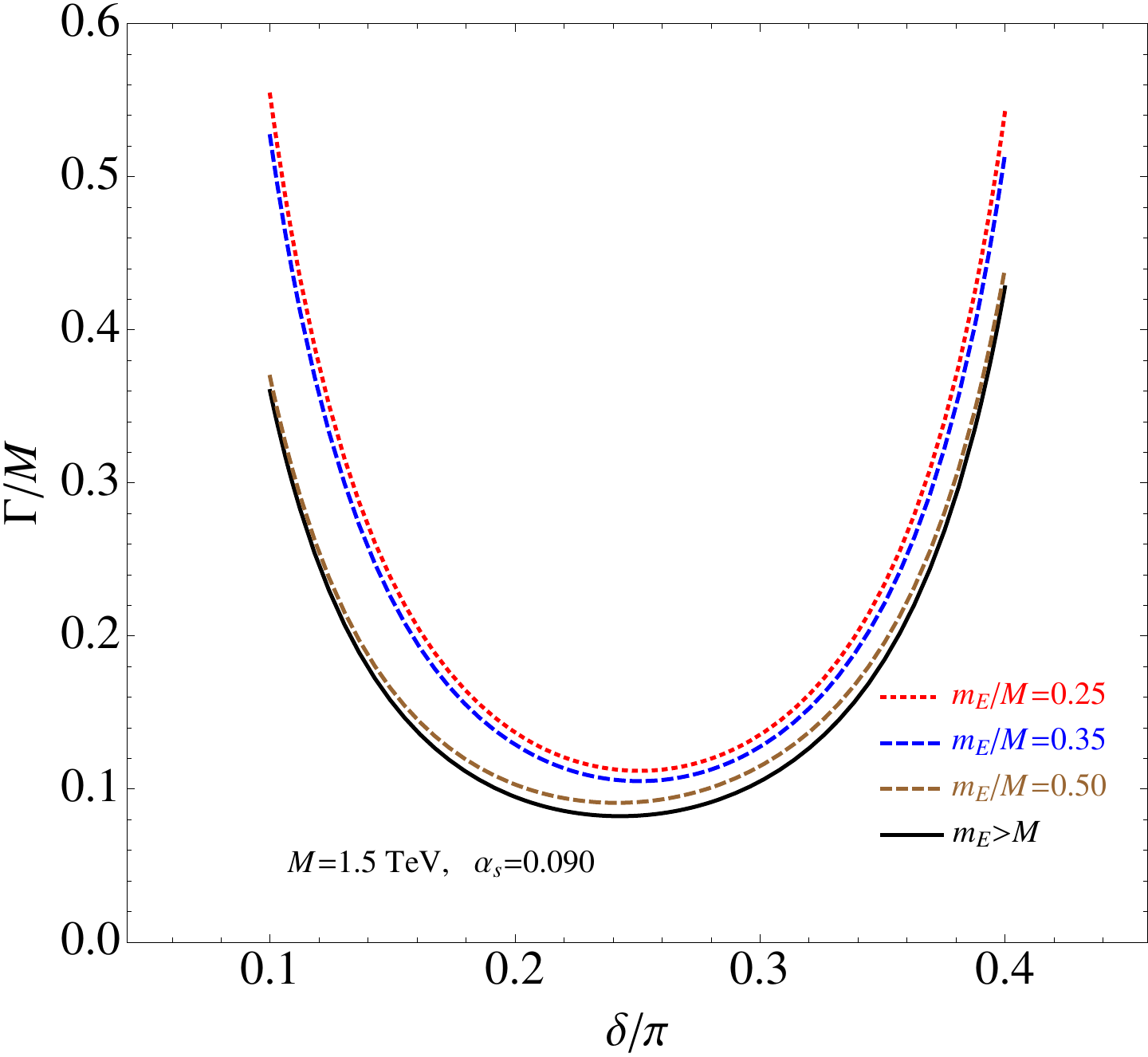}
\caption{Left: The fractional axigluon width $\Gamma/M$ for $M=1.5$ TeV and $\alpha_s (M)=0.090$ as a function of
$\delta=\tan^{-1} (g_R/g_L)$, for decays into ordinary quarks only (solid-black line at bottom), and for decays into ordinary quarks, three families of exotic pairs, and mixing-induced $t \etlc+t^c\etl$ pairs, which are shown for various common exotic masses 
 $m_E \le M/2$ as dotted-red, dashed-blue, and dashed-brown lines.
Normal-exotic mixing is ignored except for $t^c-\etlc$, for which we take $\cos^2 \theta_R = 0.25$. The $t^c-\etlc$ mixing is most noticeable in the asymmetry between $\delta$ and $\pi/2-\delta$ for larger values of $m_E$. The region $0.2 \lesssim \delta/\pi \lesssim 0.3$ is favored by the assumption  $M_S/M > 10^4 $ (Figure \ref{rge}).
Right: $\Gamma/M$ for one family of exotic pairs. In this case, $\cos \theta_R$ is the largest value consistent with the absence of a Landau pole in $\kappa_2$ up to  $M_S/M = 10^{15}$.}
\label{widthfigure}
\end{center}
\end{figure}

As mentioned in the Introduction, the various constraints from the $t\bar t$  and dijet cross sections
are relaxed considerably if the axigluon width is  large compared to the value  $\Gamma/M\sim 8-10$\%
expected for decays into ordinary quarks only\footnote{There are no tree-level decays into  two or three gluons~\cite{Cuypers:1990hb}.} with $|g_A|= g_s$ and $g_V=0$.
In particular, it has been emphasized that decays into an ordinary and exotic quark, $G_A\ra q q^c\, ' + q^c q'$
can increase the width if the off-diagonal couplings are sufficiently large~\cite{Barcelo:2011vk,Barcelo:2011wu}.
However, such couplings are proportional to  mixing angles such as  $\theta_{L,R}$ in \refl{umixing},
which we expect to be small except for the $t^c-\etlc$ mixing. Nevertheless, a large width for decays into ordinary quarks alone can be obtained
for $g_L\ne g_R$, as can be seen in Figure \ref{widthfigure}, because of the enhanced couplings in  \refl{axial}. 
Decays into exotic $q' q^c\, '$ pairs  are not suppressed by small mixings if they are kinematically allowed, and can also significantly increase the width. E.g., $\Gamma/M\sim 15$\% is obtainable for $g_L=g_R$ and a common exotic mass\footnote{One expects $m_{E_i} \sim \mathcal{M}_i$ for $i=u,d,c,s,b$, where $\mathcal{M}_i$ is the relevant singlet or doublet exotic mass in \refl{umass} or \refl{Qmass}. The exotic partner of the top quark
has $m_{E_t}\sim \sqrt{\delta_t^2+ \mathcal{M}_t^2}$.} $m_E \sim M/4$ for three exotic families, and much larger widths are
possible for $g_L\ne g_R$. Decays into scalar partners of the ordinary and exotic quarks, or into the states associated with $\Phi$ and $\ov \Phi$, could increase the width even more.

A strong constraint on or possible signature of axigluon models is associated with the production and decay of the exotic quarks. These may be pair produced through ordinary QCD processes or, if they are lighter than $M/2$,
through axigluon decays. They may also be produced singly in association with a light quark by axigluon decays or via a virtual $W$ or $Z$, but in these cases the rates are suppressed by (usually) small mixings.
Exotic decays will often be dominated by mixing with the ordinary quarks~\cite{Atre:2011ae,Barcelo:2011wu,Bini:2011zb}, leading to $q' \ra q W$, $qZ$, or 
$q H$, with the $W$, $Z$, or Higgs on-shell, for which there are already significant limits~\cite{Aad:2011yn,Aad:2012en,Aad:2012ak}. In some cases there may also be cascades involving  lighter exotics (especially in the doublet case) or scalar partners~\cite{Kang:2007ib}.
These and the $\Phi$ and $\ov \Phi$ decays   (e.g., via the couplings such as 
$\lambda_{u'} u' u^c\,' \Phi$) have rates and characteristics strongly dependent on the various masses.

The magnitudes of the axigluon couplings are universal for ordinary or mirror quarks,
i.e., for which $q_i$ and $q^c_i$ transform as $(3,1)+(1,3^\ast)$ or $(3^\ast,1)+(1,3)$:
$g_V^i= g_V$,  $g_A^i=\pm g_A$, where $g_{V,A}$ are given by \refl{axial}. However, nonuniversal magnitudes 
(or off-diagonal couplings) can be induced by mixing with exotics or other quarks with different transformations, as detailed in Appendix \ref{mixingcouplings}. Significant mixing of left-handed quarks tends to modify the  CKM matrix in violation of observations, while right-handed mixings are not so constrained. 
We have seen that in the absence of trifundamental Higgs fields it is difficult to avoid significant
top quark mixings  unless Yukawas such as $h_t$ in \refl{tmass} or $\kappa_2$ in \refl{onef} are quite large.
We therefore restrict consideration to singlet third-family exotics so that these are in the
right-handed $t^c-t^c\,'$ sector.
For the other quarks in the three exotic family case the mixings can be small. In particular, small $b$ quark mixing would lead to
an enhanced $b \bar b $ asymmetry relative to $t \bar t$. However, it is possible to have large right-handed mixings for the lighter quarks  as well, e.g., for singlet exotics
if the relevant $\delta$'s in \refl{umass}, \refl{tmass}, or their analogs for $d$, $s$, and $b$, are comparable to the $\mathcal{M}$'s. For one-exotic family some right-handed mixings must be large, but these can be restricted to the second family if desired. 
From \refl{mixdcoers} one sees that right-handed mixing (with $\theta_L$ small) has the effect of reducing the axial couplings  by $\cos^2 \theta_R$, while the vector couplings are increased or decreased depending on the sign of $\delta-\pi/4$.
Of course, $|\cos \theta_R|=1$ yields a purely vector coupling $g_s \cot \delta\ (-g_s \tan \delta)$
since the left and right mass eigenstates are both associated with $SU(3)_L$ ($SU(3)_R$).

Family-nonuniversal magnitudes can also be  obtained in models with one or more additional $SU(3)$ nodes (e.g.,~\cite{Chivukula:2010fk,Zerwekh:2011wf,Bai:2011ed}), which also involve additional massive colored vectors.
However, additional $SU(3)$ factors would enormously complicate the constructions (and also tend to decrease the Tevatron asymmetry~\cite{Chivukula:2010fk}), and will not be considered here further.

\subsection{Comparison with Collider Data}\label{signal}
\begin{figure}[htbp]
\begin{center}
\includegraphics*[scale=0.5]{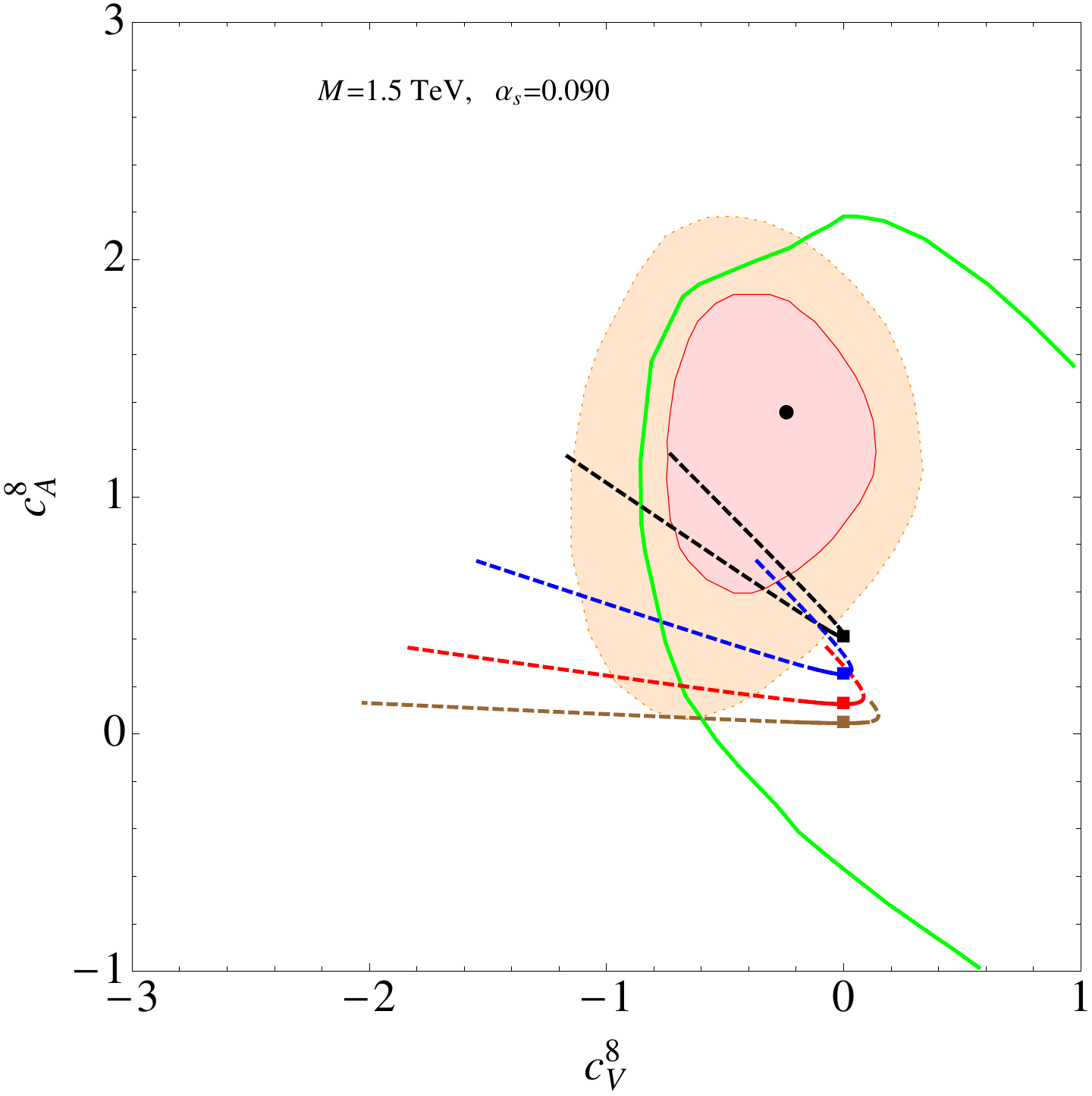}
\includegraphics*[scale=0.5]{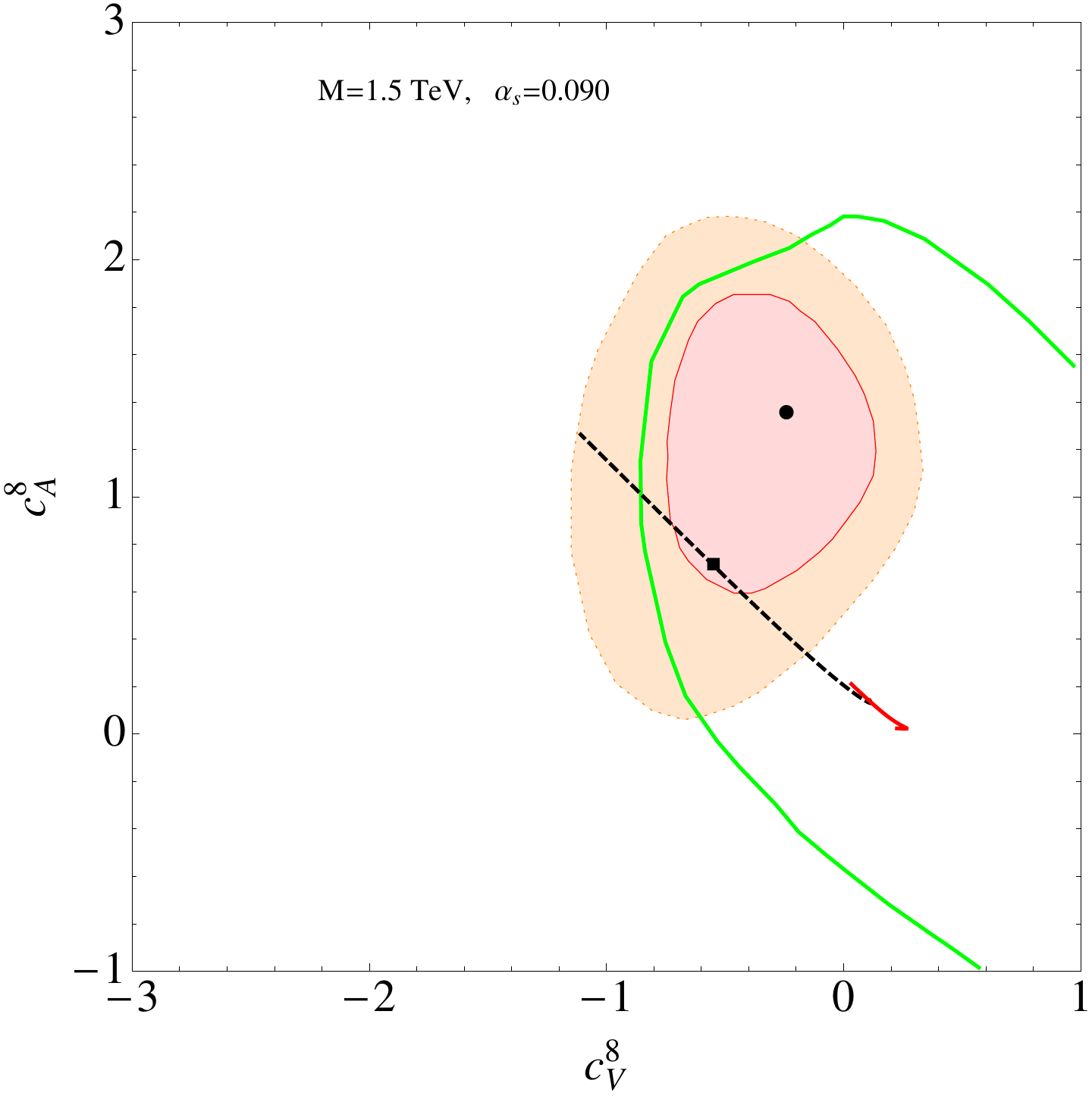}
\caption{Left: Best fit point and allowed regions in $c_V^8$ vs. $c_A^8$ at $1\sigma$ (red) and $2\sigma$ (orange) from the Tevatron $t\bar t$  cross section and asymmetry measurements, and the region excluded by the CMS $t\bar t$ cross section (exterior of green contour), reproduced from Delaunay et al~\cite{Delaunay:2011gv}.     The dashed contours are the predictions of the axigluon model for   $M=1.5$ TeV, $\alpha_s(M)=0.090$, $0.1 < \delta/\pi < 0.4$ (with 0.4 the leftmost point on each contour), and various top mixing angles $\cos \theta_R$.   From top to bottom, $\cos \theta_R=0.9$ (black), $1/\sqrt{2}$ (blue), $0.5$ (red), and $0.3$ (brown).     The favored range $0.2 < \delta/\pi < 0.3$     are the solid parts of the contours, while $\delta/\pi=0.25$ is indicated by the filled squares.  
Right: Predictions of the model with one exotic family (black dashed contour)  for $0.1 < \delta/\pi < 0.4$. For each $\delta$,
$\cos \theta_R$  is the largest value consistent with the absence of a Landau pole in $\kappa_2$ up to  $M_S/M = 10^{15}$.
The leftmost point is for $\delta/\pi=0.4$ and the black square corresponds to $\delta/\pi\sim 0.37$. 
The solid red contour also includes a $u^c-u^c_3$ mixing with $\cos \theta^u_R = 1/\sqrt{6}$.
}
\label{contours}
\end{center}
\end{figure}

Some comparison of the types of axigluon models considered here with the Tevatron and LHC data is in order. 
The collider and other implications of axigluon models have been extensively studied by other authors  (see, e.g.,~\cite{Bai:2011ed,Haisch:2011up,Delaunay:2011gv} and the other references given in the Introduction). It is not our goal to repeat those analyses. Rather, we will utilize a recent effective operator study by  Delaunay et al~\cite{Delaunay:2011gv} of the
Tevatron and LHC $t \bar t $ cross section and asymmetry measurements. 
The Tevatron data are strongly dominated by $u \bar u$ scattering, so the
 most important operators are
\beq 
\mathcal{O}_V^8= (\bar u \gamma_\mu L^i u )\, (\bar t \gamma^\mu L^i t) , \qquad
\mathcal{O}_A^8= (\bar u \gamma_\mu\gamma^5  L^i u) \, (\bar t \gamma^\mu \gamma^5 L^i t),
\eeql{effops}
with $\lag = c_V^8 \mathcal{O}_V^8 +  c_A^8 \mathcal{O}_A^8$. The best fit point and allowed regions at 1 and 2$\sigma$ from the 
Tevatron data are reproduced in Figure \ref{contours}, along with a region excluded by a recent CMS cross section measurement~\cite{Chatrchyan:2012ku}.

The coefficients $c_{V,A}^8$ for the heavy axigluon case correspond approximately to
\beq
c_V^8 \sim -\frac{g_V^u\, g_V^t}{M^2}, \qquad c_A^8 \sim -\frac{g_A^u\, g_A^t}{M^2}
\eeql{coefs}
in our notation, where it is assumed that $\hat s \ll M^2$ and that the axigluon width can be neglected\footnote{A more detailed analysis would have to take the full propagator into account.}.
The expected values are shown the nonuniversal case  for $M = 1.5$ TeV, $\alpha_s(M) = 0.09$, 
$0.1 < \delta/\pi < 0.4$, and various fixed values for the right-handed top quark mixing $\cos \theta_R$
in the left-hand plot in Figure \ref{contours}.
It is seen that the predicted values  are in good agreement for small mixings ($\cos \theta_R\sim  0.9$), and fall within the 2$\sigma$ contours  for $\cos \theta_R\gtrsim 0.3$. However, the most favorable points occur for $\delta/\pi \gtrsim 0.35$
or $\lesssim 0.25$, for which either $g_R$ or $g_L$ require a rather low string scale to avoid a Landau
pole for 3 exotic families (Figure \ref{rge}). Restricting to the  more favored range
$0.2 < \delta/\pi < 0.3$  takes one outside of the $2\sigma$ region unless $\cos \theta_R$ is close to unity.
The right-hand plot shows the predicted contour for the one exotic family case, assuming that
the top mixing  $\cos \theta_R$ for a given $\delta$ is the largest value consistent with $M_S/M = 10^{15}$.
Relatively large values of $ \delta/\pi $ fall within the $1\sigma$ contour, both because they correspond to large $g_R$
and because a large $\cos \theta_R$ (e.g., 0.9 for  $ \delta/\pi \sim 0.37$)  is allowed.

There are other collider constraints on axigluon models, including the $t \bar t$ charge asymmetry $A_C$, the dijet cross section,
and the dijet angular distribution. These are rather model dependent, but we briefly discuss the implications of the present scenario. The current charge asymmetry measurements~\cite{Chatrchyan:2011hk,ATLAS:2012an}
are consistent with the standard model. These do not yet seriously constrain the axigluon model~\cite{Delaunay:2011gv}.
However, it has been emphasized~\cite{AguilarSaavedra:2012va,Drobnak:2012cz} that better agreement would be possible
if one allowed  unequal couplings to the $u$ and $d$ because of their different relative contributions at the Tevatron and LHC.  Within the present framework $g_A^d$ could indeed be made smaller than $g_A^u$ if there were significant $d^c- d^c\,'$ mixing. However, the sign could not be reversed (as favored in~\cite{Drobnak:2012cz}) without invoking large left-handed mixing as well. Constraints from the  dijet (or $t\bar t$) cross section could be reduced for a sufficiently broad axigluon (Section \ref{width}).  However, the measured
dijet angular distribution~\cite{Aad:2011aj,Chatrchyan:2012bf} is a significant constraint
even for a large width~\cite{Haisch:2011up,Delaunay:2011gv}. One way to avoid the dijet cross section and angular distribution constraints is to assume that the axigluon couplings to the $t$ are much larger than to the $u$~\cite{Delaunay:2011gv,Barcelo:2011vk,Delaunay:2012kf}. For example, it was argued in~\cite{Delaunay:2012kf},
that a ratio $g^u/g^t \sim 1/6$ would suffice, at least for axial couplings.  Unfortunately, while it is straightforward to reduce $|g_A^u|$ by mixing in the present framework, there is no obvious way to significantly  increase $|g_A^t|$  while keeping the gauge couplings reasonably perturbative. This is illustrated for the one exotic family case in Figure \ref{contours}, 
where the effect of including a right-handed $u$-quark mixing  $\cos \theta^u_R = 1/\sqrt{6}$ (which affects both $g_V^u$ and $g_A^u$) is shown. It is seen that the effective couplings are brought close to zero, well outside the experimental region.
The analogous contours in the left-hand side of Figure  \ref{contours} are also brought very close to the origin, but are not shown for clarity.

\subsection{Field Theory Conclusions}

This Section has described some of the issues encountered in an ultraviolet-complete
field theoretic description of the types of chiral $SU(3)_L \times SU(3)_R$ axigluon model that have
been suggested to explain the Tevatron data. Although it is relatively easy to generate a
large axigluon width,  our string-inspired assumptions of no Landau poles up to a moderate or large string scale and no trifundamentals
make it  difficult but not impossible  to generate realistic quark masses and mixings without
significant  reductions of the top axial coupling due to mixing. Other serious issues 
and constraints involve FCNC, the nonobservation of exotic decays, and the difficulty of generating
family nonuniversal magnitudes of the axigluon couplings. Such constructions may still be possible, but they would require considerable  tuning or extra assumptions.

In the next section we address in more detail whether such models are
consistent with the stringy tadpole conditions and with the extra
effective global $U(1)$ symmetries found in such constructions.

\section{String Theory Construction}\label{string}

In Section \ref{fieldtheory} we saw that the string-motivated assumption
of no trifundamental Higgs has strong implications for the structure of 
Yukawa couplings. In this section we consider string-motivated augmentations
to the field theory constructions, which lead to additional constraints.
The class of gauge theories we study are common in many regions of the
string landscape, in particular weakly coupled type II orientifold
compactifications\footnote{See \cite{Blumenhagen:2005mu,Blumenhagen:2006ci} for
in-depth reviews and \cite{Cvetic:2011vz} for a brief review, including quivers.}. These gauge theories, often depicted by a quiver
diagram\footnote{A quiver is a directed graph where nodes represent
  gauge group factors and edges represent matter fields.}, generically have (at least) three
features not present in standard field theoretic constructions:
\begin{itemize}
\item Non-abelian unitary groups are realized as $U(N)$ rather than
  $SU(N)$, introducing  $U(1)$'s into the theory
  with anomalies that  are
  cancelled by appropriate Chern-Simons terms. One important
  consequence is that fields in the same representation of the
  non-anomalous symmetries
  can be quiver distinct, i.e., they
  can have different anomalous $U(1)$ charge,
  allowing one to distinguish between them and giving interesting
  family structure. This gives a mechanism for distinguishing between
  lepton and down-type Higgs doublets. Also, the anomalous $U(1)$'s impose selection
  rules on the superpotential which can forbid couplings in
  perturbation theory, though it is possible to regenerate them at
  suppressed scales by non-perturbative effects, such as D-instantons \cite{Blumenhagen:2006xt,Ibanez:2006da,Florea:2006si,Blumenhagen:2009qh}.
\item The constraints on the chiral spectrum necessary for string
  consistency include those necessary for the absence of cubic non-abelian triangle
  anomalies, but also include ``stringy" conditions that one would
  not be led to in field theory. Quivers realizing anomaly-free field
  theories such as the MSSM often violate these conditions. See 
  \cite{Cvetic:2011iq} for a recent discussion of these constraints and
  the implications for exotic matter and $Z'$ physics.
\item The string scale $M_S$ is the natural scale in the theory, and
  is typically $\cO(M_{pl}$). Therefore, fields which form a vector
  pair under all symmetries of the theory typically have a very large mass
  and decouple from low energy physics.
\end{itemize}
We will consider the ideas of Section \ref{fieldtheory} in light of
these additional ingredients and will show that they have interesting
phenomenological implications. 

Let us briefly describe the setup for the quivers\footnote{See
  \cite{Aldazabal:2000sa} for original work on D-brane quivers from
  the bottom-up perspective. For work on IIa quivers see
  \cite{Anastasopoulos:2006da,Berenstein:2006pk,Berenstein:2008xg,Cvetic:2009yh,Cvetic:2009ez,Anastasopoulos:2009mr,Cvetic:2009ng,Cvetic:2010mm,Fucito:2010dk,Cvetic:2010dz,Cvetic:2011iq}
  and for an introduction to IIa quivers, see \cite{Cvetic:2011vz}. For related work on low mass
strings in these constructions, see \cite{Anchordoqui:2008ac,Anchordoqui:2008di,Anchordoqui:2009mm,Anchordoqui:2009ja,Anchordoqui:2011ag,Anchordoqui:2011sg}.}
studied in this paper\footnote{We will see later that some aspects of
  phenomenology are better accounted for in deformations of these
  quivers involving an extra $U(1)$ node.}.  We consider four-node
quivers with $U(3)_{a_1}\times U(3)_{a_2} \times U(2)_b \times U(1)_c$
gauge symmetry.  Generically the (trace) $U(1)$ of $U(N)$ is
anomalous, and the anomalies can be cancelled via the introduction of
appropriate Chern-Simons terms. These terms appear naturally in weakly
coupled type II string theory in the Wess-Zumino contribution to the
D-brane effective action (the generalized Green-Schwarz mechanism). In
addition, the anomalous $U(1)$ gauge bosons (and sometimes the
non-anomalous ones) receive a Stuckelberg mass due to the presence of
one of the Chern-Simons terms participating in anomaly
cancellation. However, sometimes a non-anomalous linear combination
\begin{equation}
  U(1)_G = q_{a_1} \, U(1)_{a_1} + q_{a_2} \, U(1)_{a_2} + q_b \, U(1)_b + q_c \, U(1)_c
\eeql{gcharge}
remains massless. 
We demand that one
such linear combination can be identified as weak hypercharge, so that the
gauge symmetry after the Green-Schwarz mechanism lifts the anomalous
$U(1)$'s is $SU(3)_L \times SU(3)_R \times SU(2) \times U(1)_Y$. 
The presence of a field  $(3,3^*,1)_0$ to break
$SU(3)_L \times SU(3)_R$ to QCD requires $q_{a_1} = q_{a_2}$.  For
simplicity  we study only one of the two possible
hypercharge embeddings, given by $U(1)_Y = \frac{1}{6}U(3)_{a_1} +
\frac{1}{6}U(3)_{a_2} + \frac{1}{2} U(1)_c$.
The other is given by $U(1)_Y = -\frac{1}{3}U(3)_{a_1} -
\frac{1}{3}U(3)_{a_2} + \frac{1}{2} U(1)_b$.

Two sets of constraints  must be imposed. 
We follow the conventions of \cite{Cvetic:2011iq} and
refer the reader to the discussions there for more details. 
The first set of constraints are those necessary for tadpole cancellation,
which include the  cancellation of
cubic non-abelian anomalies but also include some string constraints. There
is a tadpole constraint for each node in the quiver, labeled by the ``T-charges''  $T_{a_1}$,
$T_{a_2}$, $T_b$, and $T_c$. Each matter field contributes some amount to these  T-charges. The
tadpole conditions are that the net $T_c$ charge must
be $0$ mod $3$ and the others must be $0$. There are also ``M-charge''
constraints necessary   for a massless hypercharge.
Matter fields contribute analogously to quantities $M_{a_1}$, $M_{a_2}$,
$M_{b}$, and $M_c$. 
A massless hypercharge
requires that each of the net M-charges is zero.

We take a modular approach.
First we
consider the possible quark and lepton/Higgs sectors for the quivers,
where the former give rise to the quark sector of the MSSM upon
Higgsing $SU(3)_L\times SU(3)_R$ to QCD, and the latter sector contains
fields in  the representations of the lepton and Higgs sector
of the MSSM. Any  quiver of this sort
has $SU(3)_L^3$, $SU(3)_R^3$,
$SU(3)_L^2\times U(1)_Y$ and $SU(3)_R^2 \times U(1)_Y$ triangle
anomalies. We then
consider the introduction of a colored exotic sector which cancels
these anomalies, as described in Section \ref{anomaly}, though we rely on the presence of Chern-Simons terms
to cancel the anomalies associated with the anomalous $U(1)$'s.
Such a quiver
 is anomaly free. 
However, it is possible that it does not satisfy all of the conditions necessary for string
consistency. We therefore add a colorless exotic sector for the
 purpose of satisfying these constraints and study the associated
 phenomenology.  Such a colorless exotic sector is of course possible,
 though not necessary, in field theory.

We will first consider the possible standard model quark and lepton
sectors, and then will introduce the possible exotic sectors which could
be added for anomaly cancellation. We will then show that simple phenomenological
assumptions significantly restrict the possibilities, and will study issues
of effective Yukawa couplings and  scales in two  viable quivers.
Brief comments are made on possible deformations and on matter extensions needed to satisfy the tadpole conditions.

\subsection{Classification of Quiver Sectors}
\label{classification}
In this section we classify the possible standard model quark and lepton sectors, and then
the possible exotic quark sectors. 
\subsubsection{Possible Standard Model Quark and Lepton Sectors}\label{qlsector}
The possible quark and lepton sectors
are independent of the choice of exotics. There are four
possible realizations of an ordinary generation of quarks  and four
of a mirror generation,  given by\footnote{Technically $G_2$
  and $\tilde G_2$ have $T_c = 3$, but the condition necessary
  for tadpole cancellation is $\sum_{\text{fields}} T_c^{\text{field}}
  = 0\, \text{mod} \,3$, so $T_c=3$ is effectively $0$. The same
  comment also applies to $M_2$ and $\tilde M_2$.}
\begin{center}
  \begin{tabular}{c|ccc|cccccccc}
    label & $Q$ & $u^c$ & $d^c$ & $T_{a_1}$ & $T_{a_2}$ & $T_b$ & $T_c$ & $M_{a_1}$ & $M_{a_2}$ & $M_b$ & $M_c$\\ \hline
    $G_1$& $(a_1,b)$ & $(\ov a_2, \ov c)$ & $(\ov a_2, c)$ & $2$ & $-2$ & $3$ &  $0$ &  $0$ & $0$ &$ -\frac{1}{2}$ &  $1$ \\
    $G_2$& $(a_1,b)$ & $(\ov a_2, \ov c)$ & $\Yasymm_{a_2}$ & $2$ & $-2$ & $3$ &  $0$ &  $0$ & $0$ &$ -\frac{1}{2}$ &  $1$\\
    $\tilde G_1$& $(a_1,\ov b)$ & $(\ov a_2, \ov c)$ & $(\ov a_2, c)$ & $2$ & $-2$ & $-3$ &  $0$ &  $0$ & $0$ &$ -\frac{1}{2}$ &  $1$\\
    $\tilde G_2$& $(a_1,\ov b)$ & $(\ov a_2, \ov c)$ & $\Yasymm_{a_2}$ & $2$ & $-2$ & $-3$ &  $0$ &  $0$ & $0$ &$ -\frac{1}{2}$ &  $1$
  \end{tabular}
\vspace{.5cm}

  \begin{tabular}{c|ccc|cccccccc}
    label & $Q_3$ & $t^c$ & $b^c$ & $T_{a_1}$ & $T_{a_2}$ & $T_b$ & $T_c$ & $M_{a_1}$ & $M_{a_2}$ & $M_b$ & $M_c$\\ \hline
    $M_1$& $(a_2,b)$ & $(\ov a_1, \ov c)$ & $(\ov a_1, c)$ & $-2$ & $2$ & $3$ &  $0$ &  $0$ & $0$ &$ -\frac{1}{2}$ &  $1$\\
    $M_2$& $(a_2,b)$ & $(\ov a_1, \ov c)$ & $\Yasymm_{a_1}$ & $-2$ & $2$ & $3$ &  $0$ &  $0$ & $0$ &$ -\frac{1}{2}$ &  $1$\\
    $\tilde M_1$& $(a_2, \ov b)$ & $(\ov a_1, \ov c)$ & $(\ov a_1, c)$ & $-2$ & $2$ & $-3$ &  $0$ &  $0$ & $0$ &$ -\frac{1}{2}$ &  $1$\\
    $\tilde M_2$& $(a_2, \ov b)$ & $(\ov a_1, \ov c)$ & $\Yasymm_{a_1}$ & $-2$ & $2$ & $-3$ &  $0$ &  $0$ & $0$ &$ -\frac{1}{2}$ &  $1$
  \end{tabular}
\end{center}
For example, $(a_2, b)$ is a field of representation
$(1,3,2)_Y$ with non-zero anomalous $U(1)$ charges $Q_{a_2}=Q_{b} = +1$, so that (using \refl{gcharge}) $Y = q_{a_2} + q_b=1/6$. Of course, the 2 and $2^\ast$ are equivalent under $SU(2)$, so that, e.g., $(a_2,b)$ and $(a_2, \ov b)$ differ only in their $Q_b$ charges of $+1$ or $-1$, respectively.
Young tableaux are used to denote symmetric
and antisymmetric representations, where the appropriate anomalous $U(1)$
charge is $\pm 2$ depending on whether or not the Young diagram has a bar.

The ordinary and mirror generations are equivalent  except that
$a_1\leftrightarrow a_2$. The realizations labeled with a tilde differ only by the replacement $b \rightarrow \ov b$. In the next section we will see that  
$G_2,\tilde G_2, M_2,$ and $\tilde M_2$ introduce significant difficulties
into model building, and therefore we will not consider them further.

Let us address the possible
lepton/Higgs sectors. We require\footnote{We do not consider neutrino masses here. Quiver extensions involving right-handed neutrinos are discussed in~\cite{Cvetic:2011iq}. Various possibilities for small Majorana or Dirac neutrino masses  in this context are considered in~\cite{Blumenhagen:2006xt,Ibanez:2006da,Cvetic:2008hi,Cvetic:2010mm} and reviewed in~\cite{Langacker:2011bi}.} 
the presence of $L, e^c, H_u$ and $H_d$. The differences
between sectors
amount to different choices for the $L$, $H_d$, and $H_u$, since $e^c$ can only appear as the symmetric product $\Ysymm_c$.
The three $e^c$ fields together contribute $M_c=-4$. 
The up-type Higgs  $(H_u)$  representation $(1,1,2)_{\frac{1}{2}}$ can appear as $(b,c)$ or $(\ov b,c)$, 
whereas the leptons $L$ and the down-type Higgs $(H_d)$ representation $(1,1,2)_{-\frac{1}{2}}$
can appear as $(b,\ov c)$ or $(\ov b, \ov c)$. We can therefore label
a lepton/Higgs sector as $L_{mnop}$ where $m$, $n$, $o$, and $p$ are the number of fields transforming
as $(b,c)$, $(\ov b,c)$, $(b,\ov c)$ and $(\ov b, \ov c)$, respectively. 
If we take $m+n=1$ and $o+p=4$, as in the MSSM,  their contribution
to $b$ and $c$ gives $T_b=m-n+o-p$, $M_b=\frac{3}{2}$ and $M_c=1$.
Together with the  three $e^c$'s,
 the contribution of  $L_{mnop}$ is
\begin{equation}
T_b=m-n+o-p \qquad M_b=\frac{3}{2} \qquad M_c=-3.
\end{equation}
It is  simple  to show that any given lepton/Higgs sector  with the MSSM content satisfies 
$T_b^{l/H} \in \{\pm 5,\pm 3,\pm 1\}$.

There are phenomenological advantages to choosing $(m,n,o,p)=(1,0,1,3)$ or $(0,1,3,1)$, 
in which case,  $T_b^{l/H} = -1$ or $+1$. These choices distinguish the three $L$ doublets from the $H_d$ by their
$Q_b$ charges, at least at the perturbative level. 
Furthermore,  $H_u$ and $H_d$  must have the same anomalous $Q_b$ charge to simultaneously
generate  the effective Yukawas described in Section \ref{effectiveyukawas} for both the charge $2/3$ and charge $-1/3$ quarks at the perturbative level (for either three or one exotic familes). Having the same  $Q_b$ charge also
prevents $H_u$ and $H_d$ from being a vector pair\footnote{One would still have the problem that $H_u$ and $L$ would form a vector pair, requiring some other mechanism, such as the deformation to another $U(1)$ node, to avoid a large $H_u L$ mass term.}, thus avoiding a string-scale $\mu$ parameter.
An electroweak-scale $\mu$ parameter can then be generated either by D instantons \cite{Blumenhagen:2006xt,Ibanez:2006da}
or by the VEV of a standard model singlet field~\cite{Cvetic:2010dz}.

Upon combining any quark sector 
with any lepton/Higgs sector, the resulting quiver has non-zero $T_{a_1}$ and
$T_{a_2}$ charge, corresponding to the presence of 
$SU(3)_{L,R}^3$ anomalies. In the subsections that follow, we will consider different
colored exotic sectors that can be introduced for the purpose of anomaly cancellation.

\subsubsection{Possible Exotic Quarks}

The exotic quarks of Section \ref{fieldtheory} can
be realized as
\begin{center}
\begin{tabular}{c|cccc|ccccccccc}
label & $\eulc$ & $\edlc$ & $\eul$ & $\edl$ & $T_{a_1}$ & $T_{a_2}$ & $T_b$ & $T_c$ & $M_{a_1}$ & $M_{a_2}$ & $M_b$ & $M_c$ \\ \hline
$G_E^1$ & $(\ov a_1,\ov c) $& $(\ov a_1, c)$ & $(a_2,c)$ & $(a_2,\ov c)$ & $-2$ & $2$ & $0$ & $0$ & $0$ & $0$ & $0$ & $0$ \\
$G_E^2$ & $(\ov a_1,\ov c) $& $(\ov a_1, c)$ & $(a_2,c)$ & $\ov \Yasymm_{a_2}$ & $-2$ & $2$ & $0$ & $3$ & $0$ & $0$ & $0$ & $0$ \\
$G_E^3$ & $(\ov a_1,\ov c) $& $\Yasymm_{a_1}$ & $(a_2,c)$ & $(a_2,\ov c)$ & $-2$ & $2$ & $0$ & $0$ & $0$ & $0$ & $0$ & $0$ \\
$G_E^4$ & $(\ov a_1,\ov c) $& $\Yasymm_{a_1}$ & $(a_2,c)$ & $\ov \Yasymm_{a_2}$ & $-2$ & $2$ & $0$ & $3$ & $0$ & $0$ & $0$ & $0$ \\
\end{tabular} 

\vspace{.5cm}
\begin{tabular}{c|cccc|ccccccccc}
label & $\etlc$ & $\eblc$ & $\etl$ & $\ebl$ & $T_{a_1}$ & $T_{a_2}$ & $T_b$ & $T_c$ & $M_{a_1}$ & $M_{a_2}$ & $M_b$ & $M_c$ \\ \hline
$M_E^1$ & $(\ov a_2,\ov c) $& $(\ov a_2, c)$ & $(a_1,c)$ & $(a_1,\ov c)$ & $2$ & $-2$ & $0$ & $0$ & $0$ & $0$ & $0$ & $0$ \\
$M_E^2$ & $(\ov a_2,\ov c) $& $(\ov a_2, c)$ & $(a_1,c)$ & $\ov \Yasymm_{a_1}$ & $2$ & $-2$ & $0$ & $3$ & $0$ & $0$ & $0$ & $0$ \\
$M_E^3$ & $(\ov a_2,\ov c) $& $\Yasymm_{a_2}$ & $(a_1,c)$ & $(a_1,\ov c)$ & $2$ & $-2$ & $0$ & $0$ & $0$ & $0$ & $0$ & $0$ \\
$M_E^4$ & $(\ov a_2,\ov c) $& $\Yasymm_{a_2}$ & $(a_1,c)$ & $\ov \Yasymm_{a_1}$ & $2$ & $-2$ & $0$ & $3$ & $0$ & $0$ & $0$ & $0$ \\
\end{tabular}

\vspace{.5cm}

\begin{tabular}{c|cc|ccccccccc}
label & $\tQ$ & $\tQc$ & $T_{a_1}$ & $T_{a_2}$ & $T_b$ & $T_c$ & $M_{a_1}$ & $M_{a_2}$ & $M_b$ & $M_c$ \\ \hline
$G_E^d$ & $(a_2, b) $& $(\ov a_1,\ov b)$ & $-2$ & $2$ & $0$ & $0$ & $0$ & $0$ & $0$ & $0$ \\
$\tilde G_E^d$ & $(a_2, \ov b) $& $(\ov a_1,  b)$ & $-2$ & $2$ & $0$ & $0$ & $0$ & $0$ & $0$ & $0$ \\
$M_E^d$ & $(a_1, b) $& $(\ov a_2,\ov b)$ & $2$ & $-2$ & $0$ & $0$ & $0$ & $0$ & $0$ & $0$ \\
$\tilde M_E^d$ & $(a_1, \ov b) $& $(\ov a_2, b)$ & $2$ & $-2$ & $0$ & $0$ & $0$ & $0$ & $0$ & $0$ \\
\end{tabular}
\end{center}
where $\ov \Yasymm_{a_2}$  is the antisymmetric product of two $3^\ast$'s. 
The first two tables involve exotic quarks which are singlets of $SU(2)$, and the third
involves $SU(2)$ doublet exotics (denoted by the superscript $d$). 

\subsection{Simplifying Assumptions}
\label{assumptions}
In this section we make a number of reasonable assumptions which significantly reduce the number
of viable axigluon quivers.  First consider the case of three exotic families.
\begin{enumerate}
\item The  singlet exotic generations $G_E^{2,3,4}$ and $M_E^{2,3,4}$  contain fields
  which transform as antisymmetric tensors and
  carry charge $\pm 2$ under $U(1)_{a_1}$ or $U(1)_{a_2}$. Couplings of such exotics to
  $\Phi$ or $\ov \Phi$ are forbidden by anomalous $U(1)$'s
  at the perturbative level, preventing the large exotic mass terms
  such as $\mathcal{M}_s \eul \eulc$ in \refl{umass}. We therefore we do not consider these generations. We rename
  $G_E^s\equiv G_E^1$ and $M_E^s\equiv M_E^1$ for convenience, where
  the $s$ denotes that the exotics are singlets of $SU(2)$.
  
  \item The generations $G_2, \tilde G_2, M_2, \tilde M_2$ also contain $d^c$ or $b^c$
  fields which transform as antisymmetric tensor representations. The anomalous
  $U(1)$ charge forbids perturbative couplings analogous to those in Section \ref{effectiveyukawas}
  for $u^c$ or $t^c$.
  This
  makes the generation of standard model quark Yukawa couplings as
  effective operators even more difficult, and therefore we do not
  consider these generations. We rename $G\equiv G_1$, $\tilde G \equiv \tilde G_1$,
  $M\equiv M_1$, and $\tilde M \equiv \tilde M_1$.

\item The quiver symmetries $a_1 \leftrightarrow a_2$ and
  $b\leftrightarrow \ov b$ introduce redundancies when building
  colored sectors. For example, $GGG$ and $\tilde G \tilde G \tilde G$
  are equivalent under $b\leftrightarrow \ov b$. Utilizing these
  symmetries, it is sufficient to consider $GGG, GG\tilde G, GGM,
  G\tilde GM, GG\tilde M,$ and $G\tilde G \tilde M$. \label{families} 
  
\item 
There are two ways to realize each of the standard model Higgs
  doublets: $H_u \sim (b,c)$, $\tilde H_u \sim (\ov  b,c)$, $H_d \sim
  (b, \ov c)$ and $\tilde H_d \sim (\ov b, \ov c)$, where the tilde
  denotes the sign under
 $U(1)_b$.  We will allow for the possibility
  of extensions of the MSSM spectrum involving additional Higgs pairs.
  However, we do not consider any models in which two distinct types of Higgs, e.g., $H_d$  {\em and}
  $\tilde H_d$, must both be present and have nonzero VEVs to generate all of the needed Yukawas\footnote{Similar to the discussion in Section \ref{classification} such models would involve vector pairs such as
  $H_u + \tilde H_d$. Even if one somehow avoided a string-scale  $H_u  \tilde H_d$ mass term,
  one would still require some electroweak-scale    terms to connect the sectors (in addition to the ordinary $\mu$
  terms within each sector) to generate VEVs for both.}.
  This implies that only
  one type of the fields needs  to be present, either $H_{u,d}$ or $\tilde
  H_{u,d}$. 
  That allows the pairs $ G
  G_E^s$, $ GG_E^d$, $\tilde G G_E^d$, $ M M_E^s$, $
  M M_E^d$, and $\tilde M M_E^d$ if one works with  $\tilde H_{u,d}$. Similarly, it allows
 $\tilde G G_E^s$, $ G \tilde G_E^d$,
  $\tilde G \tilde G_E^d$, $\tilde M M_E^s$, $ M \tilde M_E^d$, and $\tilde M \tilde M_E^d$
  for $  H_{u,d}$. 
  We must consider both cases since we have already utilized
  the quiver symmetry $b \leftrightarrow \ov b$ to reduce the
  number of colored sectors.

\item The observed limits on the $V_{tb}$ strongly disfavor the doublet exotics  $\tilde M_E^d$ and $M_E^d$.  The
  only remaining generation assignments are $G G_E^s$, $G
  G_E^d$, $\tilde G G_E^d$, and $M M_E^s$ for $
 \tilde H_{u,d}$. For $  H_{u,d}$ the possibilities are $\tilde G G_E^s$, $G
  \tilde G_E^d$, $\tilde G \tilde G_E^d$, and $\tilde M M_E^s$.

\item All of the remaining generations with singlet exotics, $G
  G_E^s$, $\tilde G G_E^s$, $M M_E^s$, and $\tilde M M_E^s$, have
  vector pairs such as $u' u^c$ or $t' t^c$. However, these could be
  rendered quasichiral by adding an additional $U(1)$ node, leading to
  $\delta_{s,t}$ terms.  Similarly, the generations $G G_E^d$
  and $\tilde G \tilde G_E^d$ involve vector $Q \tQc$ pairs.  Making these
  quasichiral would require an extra $U(2)$ node. That would
  enormously complicate the construction, so we will not consider such
  generations further.  On the other hand, the $Q \tQc$ pairs in $G
  \tilde G_E^d$ and $\tilde G  G_E^d$ are quasichiral already, carrying
  a net $b$ charge or $+2$ or $-2$, respectively. They could
  presumably acquire mass non-perturbatively or by coupling to singlet fields $\ov S \equiv
  \bYasymm_b$ or $ S \equiv \Yasymm_b$, respectively, leading to the
  $\delta_d$ terms. We have not attempted to construct a potential,
  but it is reasonable to assume that the VEVs would be comparable to
  or smaller than those of $\Phi$ and $\ov \Phi$ (and similarly for
  the new VEVs in the deformed models with an extra $U(1)$ node).

\item
The remaining possibilities with three exotic families are to either: (a) construct the
  three generations using combinations of $G G_E^s$, $\tilde G
  G_E^d$, and $M M_E^s$ with $\tilde H_{u,d}$; or (b) utilize the combinations of $\tilde G G_E^s$, $G  \tilde G_E^d$,
  and $\tilde M M_E^s$ with $ H_{u,d}$.  In both cases, one is subject
  to the constraints in item \# \ref{families}.
We will focus on the cases with one mirror family, which
  corresponds to a heavy axigluon. The possibilities are then: (a) $(G
  G_E^s)^2 ( M M_E^s)$, $(G G_E^s) (\tilde G G_E^d) ( M
  M_E^s)$; or (b) $(G \tilde G_E^d)^2 (\tilde M M_E^s)$, $(G \tilde G_E^d) (\tilde G
  G_E^s) (\tilde M M_E^s)$. \label{choices}
There are some phenomenological differences between the various
  possible combinations, especially concerning the off-diagonal quark
  mixing terms and possible exotic decays.  However, in all of these possibilities except
  $(G \tilde G_E^d)^2 (\tilde M M_E^s)$ the fields in $M_E^s $ and $G_E^s$ form  vector pairs,
  and this difficulty would persist for the simplest version of the deformed model with an extra 
  $U(1)$ node. We will therefore focus on the $(G \tilde G_E^d)^2 (\tilde M M_E^s)$ example,
  involving two families of doublet exotics and one mirror family of singlets.
  \end{enumerate}
  
Similar simplifications apply to the model with one exotic family. The unique case involving singlet exotics
and for which all of the couplings in \refl{onef} are allowed is $(G G_E^s)\, G M $ with $\tilde H_{u,d}$.
This can be viewed as the limit of the  $(G G_E^s)^2 ( M M_E^s)$ model after integrating out the heavy vector pairs in 
$G_E^s M_E^s$.

\subsection{Quiver with Three Families of Exotics}
\label{sec:quiver1}
In this section we look  in detail
at the quiver with a colored  and exotic  sector given by 
$GG\tilde M \tilde G_E^d \tilde G_E^d M_E^s$.
The Higgs sector is of
the  $ H_{u,d}$ type.
Requiring that $L$ and $H_d$ can be distinguished, the lepton sector is fixed up top possible extensions needed to cancel the T and M charges. The fields in
the quiver transform as
\begin{align}
  \label{eqn:quiver1}
  \text{Standard Model:} \qquad \qquad \,\,\,
  &2 \times Q \sim (a_1,b) 
  &1 \times Q_3 \sim (a_2,\ov b)   \nonumber \\
  &2 \times d^c \sim (\ov a_2,c) 
  &1 \times b^c \sim (\ov a_1,c) \nonumber \\
  &2 \times u^c \sim (\ov a_2,\ov c) 
  &1 \times t^c \sim (\ov a_1,\ov c) \nonumber \vspace{1cm}\\ 
  \vspace{1cm}
  &H_u \sim (b, c) & H_d \sim (b, \ov c) \nonumber \\
  &3\times L \sim (\ov b,\ov c) &3\times e^c \sim \Ysymm_c
  \nonumber \\
  \text{Exotics:} \qquad \qquad \,\,\,
  & 2 \times \tQ \sim (a_2, \ov b) 
  &2 \times \tQc \sim (\ov a_1,b) \nonumber \\
  & 1 \times \ttc \sim (\ov a_2, \ov c) 
  &1 \times \ttt \sim (a_1,c) \nonumber \\
  & 1 \times \tbc \sim (\ov a_2,c) 
  &1 \times \tb \sim (a_1,\ov c) \nonumber \\
  &1 \times \Phi \sim (a_1,\ov a_2)
  &1 \times \ov \Phi \sim (\ov a_1,a_2). 
\end{align}
The subscript on $Q_3$ denotes that it is the third generation
quark doublet. We will see that the structure of
anomalous $U(1)$'s affects the scales of the parameters
 $\cM_d$, $h_d$, and $\delta_d$ appearing in
\refl{Qmass}. We will change the coupling names
slightly to distinguish between up-type and down-type couplings.

The anomalous $U(1)$ symmetries and possible instanton
effects can alter the Yukawa couplings. 
The off-diagonal standard model quark Yukawa couplings $H_u Qt^c$, $H_d Qb^c$, $ H_u Q_3 u^c$,
and $H_d Q_3 d^c$ are gauge invariant under $\Ga=SU(3)_L\times SU(3)_R \times SU(2) \times U(1)_Y$.
The first two carry anomalous $U(1)_b$ charge and are forbidden in perturbation theory. 
It is possible that they are generated non-perturbatively
by D-instanton effects, however, in which case they would be naturally suppressed.
The latter two couplings are present in perturbation theory.
Depending on moduli vacuum expectation values in a string compactification, they could become
hierarchical due to worldsheet instanton effects in type IIa, for example, though they can
also be $\cO(1)$.

The standard model Yukawa couplings $H_uQu^c$, $H_d Qd^c$, $H_uQ_3t^c$ and $H_dQ_3b^c$
are not invariant under $\Ga$ and must be obtained as effective operators.
Following Section \ref{effectiveyukawas} and changing
coupling subscripts for clarity, the first two can be obtained from the $\Ga$ invariant
couplings
\begin{align}
\label{eqn:quiver1 couplings1}
h_{d} \,  H_d \tQ d^c \qquad h_{u} \,  H_u \tQ u^c \qquad \cM_Q \tQ \tQc \qquad \delta_Q Q \tQc,
\end{align}
where the first three are present in perturbation theory and the last carries anomalous $U(1)_b$
charge and must be generated non-perturbatively. The scale of $M_Q$ is set by $v_\phi\sim \cO(TeV)$ and
therefore these fields are present in the low energy spectrum. It is convenient that $\delta_Q$ 
is naturally suppressed, however, since otherwise it would acquire a string scale mass and
decouple. The Yukawa couplings $h_d$ and $h_u$ can be $\cO(1)$ or smaller.

The third generation standard model Yukawa couplings $H_u Q_3 t^c$ and $H_d Q_3 b^c$  can be obtained as effective operators from the $\Ga$
invariant couplings
\begin{align}
h_b \, H_dQ_3\tbc \qquad \cM_b \tb \tbc \qquad \delta_b b^c \tb \nonumber \\
h_t \, H_uQ_3\ttc \qquad \cM_t \ttt \ttc \qquad \delta_t t^c \ttt. 
\label{eqn:quiver1 couplings2}
\end{align}
$h_b$ and $h_t$ are allowed in perturbation theory and can therefore be $\cO(1)$.
The masses $\cM_b$ and $\cM_t$ are set by $v_\phi$.
However, $\delta_b$ and $\delta_t$ are present
in perturbation theory  (i.e.,  $b^c \tb$ and $t^c \ttt$ are vector pairs) and therefore are generically
of order the string scale, so that the associated fields decouple. This is a significant drawback,
which we will attempt to remedy.

In addition, the pairs $(u^c, \ttc)$, $(d^c, \tbc)$, and $(\tQ, Q_3)$ have the same quantum numbers.
Since the couplings $u^c \ttt$, $d^c \tb$  and $Q_3 \tQc$ are linear in the redundant
fields, a field redefinition can rotate them away. This is not true of $H_u Q_3 u^c$, $H_d Q_3 d^c$, 
 $H_u \tQ \ttc$, and  $H_d \tQ \tbc$, however, and they can lead to small CKM mixing of the third family with the
first two. Finally, the $\mu$-term is perturbatively forbidden and can be generated
at a suppressed scale by D-instantons~\cite{Blumenhagen:2006xt,Ibanez:2006da}.

The problem of the string-scale $\delta_{b,t}$ can be fixed via a simple deformation of the above
quiver involving an additional  $U(1)_d$ node with hypercharge 
\begin{equation}
  U(1)_Y = \frac{1}{6} U(1)_{a_1}+\frac{1}{6} U(1)_{a_2} + \frac{1}{2}U(1)_c + \frac{1}{2}U(1)_d. 
  \label{4hyper}
\end{equation}
Let the singlet quark exotics and Higgs doublets in \refl{eqn:quiver1} transform under $U(1)_d$
rather than $U(1)_c$, by replacing $c$ with $d$ and $\ov c$ with $\ov d$. 
The representation theory of the quiver with respect to $\Ga$ has not changed, but the anomalous
$U(1)$ charges of the fields have changed. The couplings $h_d$, $h_u$,
$\delta_b$, and $\delta_t$ are now forbidden in perturbation theory because they are  protected by a symmetry, but small values such as $\delta_t = \mathcal{O}(v_\phi)$ can still be generated by D-instantons.  The same applies to the $\Ga$-invariant standard model Yukawa couplings, which now carry
anomalous $U(1)$ charge. 
The
redundancy between $Q_3$ and $\tQ$ is still present, but not  between $u^c$ and $\ttc$
or $d^c$ and $\tbc$.
$Q_3 \tQc$ can still be rotated away, while  $H_u \tQ \ttc$ and  $H_d \tQ \tbc$  (and
$H_uQ_3t^c$ or $H_dQ_3b^c$  if they are generated by D instantons) can still lead to small CKM mixings.

Let us discuss the field content. While this quiver is consistent as a quantum field theory
 (with the addition of Chern-Simons terms to cancel abelian and/or mixed anomalies)
it does not satisfy the conditions necessary for tadpole cancellation. Specifically, all
of the T-charge and M-charge conditions are satisfied except for $T_b=2$, when for consistency
it must be zero. There are only two  ways that matter could be added for the sake of 
consistency without overshooting the tadpole in the other direction. The first is to add two
fields transforming as $\ov b$, which in this case means a quasichiral pair of fields
$(\ov b, c)$ and $(\ov b, \ov c)$. These additions have the quantum numbers of a pair of  lepton
doublets which are vector like with respect to $\Ga$~\cite{Cvetic:2011iq}. The second possibility is to add a single field $\Yasymm_b$. This singlet
couples to $LQ_3d^c$ and $LLe_L^c$ in perturbation theory and therefore could dynamically give rise to $R$-parity violating operators.

Finally, the mass term $\Phi \ov \Phi$ of the axigluon Higgs fields will never become
charged under an anomalous $U(1)$ symmetry, and therefore these fields naturally have a high
mass. In particular, the vector pair is not lifted by the deformation we have performed, and there is
no deformation which will do so while retaining the same $\Ga$ representations for $\Phi$ and $\ov \Phi$.
One option would be to add additional $SU(3)$ factors to the gauge group, but this would significantly
complicate the construction. The mass of $\Phi \ov \Phi$ is a significant drawback of these constructions.

\subsection{Quiver with One Exotic Family}

\label{sec:quiver2}
In this section we look  in detail
at the quiver with a colored  and exotic  sector given by $GGMG_E^s$. This quiver only has one family of exotics, as discussed in Section \ref{effectiveyukawas}, and can be viewed as the limit of $(G G_E^s)^2 ( M M_E^s)$  after decoupling the vector pairs in $G_E^s M_E^s$.
 The Higgs sector is of
 the  $\tilde H_{u,d}$ type, and we again require that $L$ and $H_d$ can be distinguished. Up to possible extensions needed to cancel the T and M charges, the fields are 
 \begin{align}
  \label{eqn:quiver2}
  \text{Standard Model:} \qquad \qquad \,\,\,
  &2 \times Q_i \sim (a_1,b) 
  &1 \times Q_3 \sim (a_2,b)   \nonumber \\
  &2 \times d_i^c \sim (\ov a_2,c) 
  &1 \times d_3^c \sim (\ov a_1,c) \nonumber \\
  &2 \times u_i^c \sim (\ov a_2,\ov c) 
  &1 \times u_3^c \sim (\ov a_1,\ov c) \nonumber \vspace{1cm}\\ 
  \vspace{1cm}
  &H_u \sim (\ov b, c) & H_d \sim (\ov b, \ov c) \nonumber \\
  &3\times L \sim (b,\ov c) &3\times e^c \sim \Ysymm_c
  \nonumber \\
  \text{Exotics:} \qquad \qquad \,\,\,
  & 1 \times \tuc \sim (\ov a_1, \ov c) 
  &1 \times \tu \sim (a_2,c) \nonumber \\
  & 1 \times \tdc \sim (\ov a_1,c) 
  &1 \times \td \sim (a_2,\ov c) \nonumber \\
  &1 \times \Phi \sim (a_1,\ov a_2)
  &1 \times \ov \Phi \sim (\ov a_1,a_2). 
\end{align}
The anomalous $U(1)$'s and possible D instantons  affect the scales of the parameters
$\cM$, $h_i$, $k_i$, $\kappa_i$, and $\delta_i$ appearing in
\refl{onef} and therefore the quark masses and mixing angles. From Section \ref{effectiveyukawas}
 the relevant mass terms are
\begin{align}
\label{eqn:quiver2 exotic couplings}
\mathcal{M}_u \eul \eulc, \,\,\,\,\,  h_{u_i} H_u Q_i u_3^c, \,\,\,\,\,  k_{u_i}  H_u Q_i  \eulc, \,\,\,\,\,  
\kappa_{u_i} H_u Q_3 u_i^c, \,\,\,\,\,  \,\,\,\,\,   \delta_{u_i} \eul  u^c_i \nonumber \\
\mathcal{M}_d \edl \edlc, \,\,\,\,\,  h_{d_i} H_d Q_i d_3^c, \,\,\,\,\,  k_{d_i}  H_d Q_i  \edlc, \,\,\,\,\,  
\kappa_{d_i} H_d Q_3 d_i^c, \,\,\,\,\,  \,\,\,\,\,   \delta_{d_i} \edl  d^c_i,
\end{align}
where we have changed the subscripts slightly for clarity.
All of these couplings are uncharged under the anomalous $U(1)$  and are present in
perturbation theory. The couplings of $h$-type, $k$-type, and $\kappa$-type can be hierarchical due
to worldsheet instanton effects in type IIa string theory, for example, though they can
also be $\cO(1)$. The scales of $\cM_d$ and $\cM_u$ are set
by $v_\phi$.
However, the $\delta$-type  mixing terms  are string scale and the associated fields can be
integrated out at low energies. An analogous problem existed for some $\delta$-terms
in Section \ref{sec:quiver1}.
Note that the fields $u_3^c$ and $\tuc$, and also $d_3^c$ and $\tdc$, have identical
quantum numbers. The couplings we have discussed are linear in these redundant fields,
however, so it is possible define a linear combination which rotates some of the terms away.
We will not discuss this in detail, since a deformation can lift the redundancy.

We can again perform a simple deformation 
to avoid string-scale masses by adding an additional  $U(1)_d$  node with the hypercharge 
given by \refl{4hyper}.
We again choose the singlet quark exotics in \refl{eqn:quiver2} to transform under $U(1)_d$
rather than $U(1)_c$. The masses $\delta_{d_i}$ and $\delta_{u_i}$ 
are now forbidden in perturbation theory but can be generated by D-instantons. 
In this case it is phenomenologically preferable to continue to associate the Higgs doublets with 
 $U(1)_c$, i.e., $H_u = (\ov b, c)$, $H_d = (\ov b, \ov c)$. Then, the $\kappa_{u_i}$ and $\kappa_{d_i}$
 terms are allowed perturbatively. Similar to Section  \ref{effectiveyukawas},
 the $\mathcal{M}$, $\delta$, and $\kappa$ terms can therefore be of the magnitude required to generate
 the $t$ quark mass. The $h$ terms are still allowed perturbatively, while the $k$ terms are non-perturbative and suppressed. For suitable values it is possible to generate the lighter quark masses and CKM mixings.

All of the T-charge and M-charge conditions are satisfied except for $T_b=10$. One possibility for solving this overshooting  is to
add five singlets transforming as $\Yasymm_b$. These fields couple to $H_uH_d$ in perturbation
theory and therefore could give rise to a dynamical $\mu$-term~\cite{Cvetic:2010dz}.
One could instead add five additional Higgs doublet pairs, or some combination of these with each other or with an $SU(2)$ triplet $\ov \Ysymm_{b}$.

The problem of the string-scale  $\Phi \ov \Phi$ mass term is identical to that in Section \ref{sec:quiver1}.

\section{Discussion}

The CDF and D0 collaborations at Fermilab have reported an
 anomalously large forward-backward asymmetry in 
$t \bar{t}$ production. One possible explanation involves the extension of QCD to
a chiral $SU(3)_L \times SU(3)_R$ color group, which is spontaneously broken to the diagonal subgroup. 
In addition to the eight massless gluons, there are an octet of massive axigluons which have axial vector (and possibly also vector) couplings to the  quarks. The axigluons may be relatively light ($\mathcal{O}(100)$
GeV), with universal couplings to all three families, or may be in the TeV range, with the $SU(3)_L$
and $SU(3)_R$ couplings reversed for the (mirror) third family.

A number of studies have indicated that the Tevatron results can be accomodated in the axigluon framework,
though there are considerable constraints and tension from other Tevatron and LHC results, including the
$t \bar t$ cross section, the dijet cross section and shapes, and the LHC $t \bar t$ charge asymmetry.
There are other constraints from flavor changing neutral currents, electroweak precision tests, etc.

Most of the existing studies have allowed arbitrary axigluon couplings, or have been in the framework of
effective field theories. In this paper we have emphasized possible ultraviolet completions of the chiral color models, especially of the type motivated by classes of perturbative superstring constructions such as
local type IIa intersecting brane models. We have found that such completions lead to many additional complications, constraints,  difficulties, and possible experimental signals not revealed at the effective field
theory level. 

Much of our analysis was done at the field theory level, but including two additional string-motivated assumptions: (a) that all gauge and Yukawa interactions remain perturbative up to a
string, compactfication, or GUT scale much larger than the TeV scale. (b) That all fields in the low energy theory
transform as bifundamentals, singlets, adjoints, or symmetric or antisymmetric products, i.e., that they cannot be simultaneously charged under three gauge factors (trifundamentals). The latter condition is motivated by a large class of string vacua.

The extension of QCD to  chiral $SU(3)_L \times SU(3)_R$ introduces
$SU(3)_L^3$, $SU(3)_R^3$, $SU(3)_L^2\x U(1)_Y$, and $SU(3)_R^2\x
U(1)_Y$ triangle anomalies, requiring the addition of exotic fields to cancel them. The form of these exotics is strongly restricted by our assumption that the Yukawa and gauge couplings  remain perturbative up to a large scale. In particular, $SU(2)$-chiral states, such as have  been considered in most earlier studies, would require large Higgs Yukawa couplings and therefore Landau poles at low scales. This suggests instead that the exotics are quasi-chiral, i.e., non-chiral with respect to the SM gauge group.
The perturbativity of the gauge couplings essentially restricts the possibilities
to either three or one family of heavy exotic quarks. Each family  consists of two
quarks, whose left and right-handed components are both 
$SU(2)$ singlets or are both combined in an $SU(2)$ doublet, and with mass associated with the chiral 
$SU(3)_L \times SU(3)_R$ breaking scale. For three exotic families the finiteness of the gauge couplings requires a relatively low string scale, no higher than $\sim 10^{11}$ GeV. For one exotic family, which is only possible for a mirror third generation, the couplings are asymptotically free, allowing a string scale as large as the Planck scale.

The absence of trifundamental Higgs fields implies that the quark Yukawa couplings must be due to higher-dimensional operators, which we assume are generated by mixing with the heavy exotic states. This is challenging for the top quark, for which the effective Yukawa coupling is of $\mathcal{O}(1)$,
and suggests significant mixing of the right-handed top  (for singlet exotics), or of the left-handed top (for doublets), with the latter case excluded by CKM matrix observations.

There are many implications of these features, including the possibility of a large axigluon width, the production and decays of the exotic quarks, nonuniversal magnitudes of the axigluon couplings due to mixing with exotics (which dilutes the $t\bar t$ asymmetry), the $b\bar b$ asymmetry, CKM universality and observations, and FCNC,
some of which can help evade other experimental searches and some of which lead to other constraints.
Our conclusion is that it is extremely difficult but  not impossible to accomodate the existing data within the framework of type II string-motivated field theory.

We also considered the embedding of the field theory models in a class of string constructions such as type IIa intersecting brane theories, making use of a quiver analysis which captures the local constraints. Such theories usually involve $U(N)$ gauge factors. The extra (trace) $U(1)$'s are typically anomalous, with the  associated gauge bosons 
acquiring string scale masses. These $U(1)$'s remain as global symmetries of the low energy theory at the perturbative level\footnote{There may be additional implications for string axions~\cite{Berenstein:2012eg} or for the strong $CP$ problem~\cite{Hsu:2004mf}.}, though they may be broken by suppressed non-perturbative effects such as D instantons.
There are also tadpole cancellation conditions, which include triangle anomalies but can be stronger, as well as conditions for a linear combination of $U(1)$'s to  be non-anomalous and correspond to hypercharge.

Although there are many possible quivers, simple phenomenololgical considerations reduce the possibilities
considerably. We examined in some detail two of these, one involving three exotic families and  one with a single exotic family. It was shown that the various couplings needed to generate exotic masses, ordinary quark masses, and CKM mixings were indeed present at either the perturbative or non-perturbative level. However, each case involved some vector pairs of fields that would be expected to acquire string-scale masses. This difficulty could be remedied, however, by deforming the quiver to contain an additional   $U(1)$ node. In each example, additional fields (typically SM singlets or  $SU(2)$-doublet pairs corresponding to additional Higgs pairs or quasichiral leptons) were required to satisfy the stringy tadpole conditions.
One serious problem with these constructions is that the chiral supermultiplets needed to break the chiral 
$SU(3)_L \times SU(3)_R$ and to generate ordinary and exotic quark masses occur as a vector pair
which cannot be prevented from acquiring a 
 string-scale mass by any simple deformation. As in the field theory case, we conclude that it would be quite difficult to explain the $t\bar t$ asymmetry and other data within this theoretical framework, but we cannot completely exclude it.
 
 Our primary motivation in this study was to examine and illustrate the additional difficulties and implications of embedding a relatively straightforward extension of the SM or MSSM in a class of top-down string or string-motivated constructions, independent of whether the Tevatron anomaly survives as new physics.

\acknowledgments

J.H. thanks Denis Klevers and Hernan Piragua for useful conversations. J.H. was supported by 
a U.S. Department of
Energy Graduate Fellowship for most of this work.
This material is based upon work supported in part by the National Science Foundation under 
Grants No. 1066293, PHY11-2591, and RTG DMS Grant 0636606, the U.S. Department of Energy under Grant DOE-EY-76-02-3071, and the hospitality of the Aspen Center for Physics. 
MC is also supported by the Fay R. and Eugene L. Langberg 
Endowed Chair and the Slovenian
Research Agency (ARRS).

\appendix
\section{Renormalization Group Equations}  \label{rgeappendix}
Above the $SU(3)_L \times SU(3)_R$ symmetry breaking scale, the $g_L$ running is described  by~\cite{Jones:1981we,Einhorn:1981sx}
\beq
\frac{d g_L}{d t} = \frac{1}{16\pi^2} \,  \left[ \frac{N_t}{2} -9 \right]\, g_L^3
+  \frac{1}{(16\pi^2)^2}   \,\left[ \frac{17 N_t}{3} -54 \right]\,  g_L^5 
+  \frac{1}{(16\pi^2)^2}   \,  \left[ 8 N_\Phi \right]\, g_L^3\, g_R^2 ,
\eeql{gaugerge}
where $t= \ln\, \mu$, $N_t=N_f + 3 N_\Phi$, $N_f$ is the number of chiral supermultiplets transforming as $(3,1)$ or $(3^\ast,1)$,
and $N_\Phi$ is the number transforming as $(3,3^\ast)$ or $(3^\ast,3)$. In the model with three (one)
families of exotics as well as $\Phi$ and $\ov \Phi$ one has $N_f=12\, (8)$ and $N_\Phi=2$,
so that the one loop coefficient is $0\, (-2)$.
We have ignored the two-loop effects of the electroweak and Yukawa couplings.
An analogous equation holds for $g_R$. The implications of  a vanishing one-loop $\beta$ function are discussed in Section \ref{anomaly}.

The one-loop RGE for the top-Yukawa coupling $y_t=m_t/v_u$ in the MSSM is (e.g.~\cite{Martin:1993zk})
\beq
\frac{d y_t}{d t} = \frac{y_t}{16\pi^2} \,  \left[6 y_t^2 - \frac{16}{3} \, g_s^2\right],
\eeql{YukawaMSSM}
where the electroweak couplings and smaller Yukawas are neglected.
It is well known (e.g.,~\cite{Dedes:2000jp}) that the absence of a Landau pole in $y_t$ up to a
scale $M_X$ places an upper bound on $y_t (M_Z)$ (and therefore a 
lower bound on   $\tan \beta\equiv v_u/v_d$).
This bound  can be approximated by the quasi fixed point value (e.g.,~\cite{Lanzagorta:1995gp})
\beq
\left. y_t (M_Z) \right|_{\text{QFP}} = g_s (M_Z) \left[\frac{7/18}{1-\left( \frac{\alpha_s (M_X)}{\alpha_s (M_Z)} \right)^{7/9}}\right]^{1/2},
\eeql{qfpy}
where $M_X$ is the scale at which $y_t \gg g_s$. Choosing $M_X \sim 2 \times 10^{16}$ yields $\left. y_t\, (M_Z) \right|_{\text{QFP}}\sim 1.02$,
which is relaxed to $\sim 1.09$ when the electroweak couplings are included.

A similar bound applies to the coupling $\kappa_2 $ in  \refl{onef} for the case of one exotic family. The RGE is
\beq
\frac{d \kappa_2}{d t} = \frac{ \kappa_2}{16\pi^2} \,  \left[6  \kappa_2^2 - \frac{16}{3}  g_R^2 \right].
\eeql{Yukawarge}
If one  uses the one-loop approximation to the $g_R$ analog of  \refl{gaugerge},
the quasi fixed point value for $ \kappa_2$ is
\beq
\left. \kappa_2 (M) \right|_{\text{QFP}} = g_R (M) \left[\frac{5/9}{1-\left( \frac{\alpha_R (M_S)}{\alpha_R (M)} \right)^{5/3}}\right]^{1/2},
\eeql{qfph}
where $M$ is the axigluon mass and $M_S$ is the string scale. Taking $M_S/M\sim 10^{15}$ and $\alpha_s (M)\sim 0.09$, 
 the QFP value increases from $\sim 1.0$ to 1.6 as $\delta/\pi$ increases from 0.1 to 0.4,
corresponding to $\kappa_2  v_u \lesssim $  170 to 450 GeV.

As discussed in Section \ref{anomaly} the model with three exotic families is only consistent for a rather low string scale.
Moreover,  one cannot use the quasi fixed point approximation for $h_t$ in \refl{tmass} because of the vanishing of the one-loop $\beta$ function.
The  upper limits on $h_t$ are therefore
difficult to calculate  without a full two-loop calculation, which is beyond the scope of this paper. For definiteness, we will consider 
the same range, 170 GeV $\lesssim h_t v_u \lesssim$ 450 GeV, as for a single exotic family.

\section{Exotic Mixing and Couplings} \label{mixingcouplings}
Consider an ordinary left-chiral quark $u_L$ transforming as $(3,1)$ under $SU(3)_L \times SU(3)_R$,
and its right-handed partner $u_R$ transforming as $(1,3)$ (i.e, $u_R$ is  the $CP$ conjugate of $u^c_L$),
 as well as an exotic mirror pair $u'_L$ and $u'_R$ transforming as $(1,3)$ and $(3,1)$.
 These contribute to the  currents $J_{L,R}^\mu$ in \refl{gauge2} as
 \beq
 J_L^\mu = \bar u_L \gamma^\mu u_L + \bar u'_R \gamma^\mu u'_R, \qquad J_R^\mu = \bar u_R \gamma^\mu u_R + \bar u'_L \gamma^\mu u'_L,
 \eeql{ordcurrent}
 where the color factors are suppressed. For a mirror family, the expressions for $J_L$ and $J_R$ are reversed.
 As discussed in Section \ref{effectiveyukawas}, in the absence of trifundamental Higgs representations the quark masses must be generated by mixing with exotics, which in turn modifies the diagonal couplings to axigluons and  induces off-diagonal ones. In the presence of mixing the original fields are related to the mass eigenstates $u_{iL}$ and $u_{iR}$, $i=1,2$, by 
 \beq
 \begin{pmatrix} u_L  \\  u'_L \end{pmatrix} = 
 \begin{pmatrix}
\, c_L & s_L  \\ -s_L & c_L 
\end{pmatrix}
\begin{pmatrix} u_{1L}  \\  u_{2L} \end{pmatrix} , \qquad 
\begin{pmatrix} u_R  \\  u'_R \end{pmatrix} = 
 \begin{pmatrix}
\, c_R & s_R  \\ -s_R & c_R 
\end{pmatrix}
\begin{pmatrix} u_{1R}  \\  u_{2R} \end{pmatrix} , 
 \eeql{mixmatrix}
 where $c_L \equiv \cos \theta_L$, etc. In terms of the mass eigenstates, the axigluon current becomes
 \beq
g_s \left( \cot \delta \, J_L^\mu - \tan \delta\,   J_R^\mu\right)  = \sum_{i,j=1,2} \bar u_i \gamma^\mu \left(g_V^{ij} - g_A^{ij} \gamma^5\right) u_j,
 \eeql{mixcurrent}
where   
\begin{align}
g_V^{11} & = g_V + g_A (s_R^2-s_L^2), \qquad  & g_A^{11}  &= +g_A ( c_L^2-1 + c_R^2)\notag \\
g_V^{22} & = g_V - g_A (s_R^2-s_L^2), \qquad & g_A^{22} & = -g_A ( c_L^2-1 + c_R^2) \label{mixdcoers} \\
g_V^{12}&=g_V^{21}= g_A (c_L s_L - c_R s_R), \qquad & g_A^{12}& =g_A^{21}= g_A (c_L s_L + c_R s_R),\notag
\end{align}
and $g_{V,A}$ are defined in \refl{axial}.
For a mirror family, $c_L \leftrightarrow c_R$, $s_L \leftrightarrow s_R$, and $g_A^{ij} \rightarrow -g_A^{ij}$.

\clearpage
\section{Table of Quiver Representations}
In table \ref{table:quiver reps} we present all possible field representations for the four-node
quivers considered in this paper. T-charges and M-charges for quivers with an additional $U(1)_d$
node can be determined by mapping $c$ to $d$ and $\ov c$ to $\ov d$ for the relevant fields.
 The
T-charges and M-charges of bifundamentals of $U(1)_c$ and $U(1)_d$ can be found in table 16 of 
\cite{Cvetic:2011iq}.

\begin{table}[htbp]
\centering
\scalebox{.7}{
\begin{tabular}[h]{c|c|c|c|c|c|c|c|c|}
  Transformation & $T_{a_1}$ & $T_{a_2}$ & $T_b$ & $T_c$ & $M_{a_1}$ & $M_{a_2}$ & $M_b$ & $M_c$ \\ \hline 
  $(\ov{a_1},b)$\qquad $(\thrb,\textbf{1},\textbf{2})_{-\frac{1}{6}}$ &$-2$ &$0$ &$3$ &$0$ &$0$ &$0$ &$\frac{1}{2}$ &$0$\\ \hline
  $(a_1,\ov{b})$\qquad $(\thr,\textbf{1},\textbf{2})_{\frac{1}{6}}$ &$2$ &$0$ &$-3$ &$0$ &$0$ &$0$ &$-\frac{1}{2}$ &$0$\\ \hline
  $(a_1,b)$\qquad $(\thr,\textbf{1},\textbf{2})_{\frac{1}{6}}$ &$2$ &$0$ &$3$ &$0$ &$0$ &$0$ &$-\frac{1}{2}$ &$0$\\ \hline
  $(\ov{a_1},\ov{b})$\qquad $(\thrb,\textbf{1},\textbf{2})_{-\frac{1}{6}}$ &$-2$ &$0$ &$-3$ &$0$ &$0$ &$0$ &$\frac{1}{2}$ &$0$\\ \hline
  $(\ov{a_2},b)$\qquad $(\textbf{1},\thrb,\textbf{2})_{-\frac{1}{6}}$ &$0$ &$-2$ &$3$ &$0$ &$0$ &$0$ &$\frac{1}{2}$ &$0$\\ \hline
  $(a_2,\ov{b})$\qquad $(\textbf{1},\thr,\textbf{2})_{\frac{1}{6}}$ &$0$ &$2$ &$-3$ &$0$ &$0$ &$0$ &$-\frac{1}{2}$ &$0$\\ \hline
  $(a_2,b)$\qquad $(\textbf{1},\thr,\textbf{2})_{\frac{1}{6}}$ &$0$ &$2$ &$3$ &$0$ &$0$ &$0$ &$-\frac{1}{2}$ &$0$\\ \hline
  $(\ov{a_2},\ov{b})$\qquad $(\textbf{1},\thrb,\textbf{2})_{-\frac{1}{6}}$ &$0$ &$-2$ &$-3$ &$0$ &$0$ &$0$ &$\frac{1}{2}$ &$0$\\ \hline
  $(\ov{a_1},c)$\qquad $(\thrb, \textbf{1},\textbf{1})_{\frac{1}{3}}$ &$-1$ &$0$ &$0$ &$3$ &$-\frac{1}{2}$ &$0$ &$0$ &$0$\\ \hline
  $(a_1,\ov{c})$\qquad $( \thr, \textbf{1},\textbf{1})_{-\frac{1}{3}}$ &$1$ &$0$ &$0$ &$-3$ &$\frac{1}{2}$ &$0$ &$0$ &$0$\\ \hline
  $(a_1,c)$\qquad $(\thr,\textbf{1},\textbf{1})_{\frac{2}{3}}$ &$1$ &$0$ &$0$ &$3$ &$-\frac{1}{2}$ &$0$ &$0$ &$-1$\\ \hline
  $(\ov{a_1},\ov{c})$\qquad $(\thrb,\textbf{1},\textbf{1})_{-\frac{2}{3}}$ &$-1$ &$0$ &$0$ &$-3$ &$\frac{1}{2}$ &$0$ &$0$ &$1$\\ \hline
  $(\ov{a_2},c)$\qquad $(\textbf{1},\thrb,\textbf{1})_{\frac{1}{3}}$ &$0$ &$-1$ &$0$ &$3$ &$0$ &$-\frac{1}{2}$ &$0$ &$0$\\ \hline
  $(a_2,\ov{c})$\qquad $(\textbf{1},\thr,\textbf{1})_{-\frac{1}{3}}$ &$0$ &$1$ &$0$ &$-3$ &$0$ &$\frac{1}{2}$ &$0$ &$0$\\ \hline
  $(a_2,c)$\qquad $(\textbf{1},\thr,\textbf{1})_{\frac{2}{3}}$ &$0$ &$1$ &$0$ &$3$ &$0$ &$-\frac{1}{2}$ &$0$ &$-1$\\ \hline
  $(\ov{a_2},\ov{c})$\qquad $(\textbf{1},\thrb,\textbf{1})_{-\frac{2}{3}}$ &$0$ &$-1$ &$0$ &$-3$ &$0$ &$\frac{1}{2}$ &$0$ &$1$\\ \hline
  $\Ysymm_{a_1}$\qquad $(\six,\textbf{1},\textbf{1})_{\frac{1}{3}}$ &$7$ &$0$ &$0$ &$0$ &$-\frac{1}{2}$ &$0$ &$0$ &$0$\\ \hline
  $\ov \Ysymm_{a_1}$\qquad $(\sixb,\textbf{1},\textbf{1})_{-\frac{1}{3}}$ &$-7$ &$0$ &$0$ &$0$ &$\frac{1}{2}$ &$0$ &$0$ &$0$\\ \hline
  $\Yasymm_{a_1}$\qquad $(\thrb,\textbf{1},\textbf{1})_{\frac{1}{3}}$ &$-1$ &$0$ &$0$ &$0$ &$-\frac{1}{2}$ &$0$ &$0$ &$0$\\ \hline
  $\ov \Yasymm_{a_1}$\qquad $(\thr,\textbf{1},\textbf{1})_{-\frac{1}{3}}$ &$1$ &$0$ &$0$ &$0$ &$\frac{1}{2}$ &$0$ &$0$ &$0$\\ \hline
  $\Ysymm_{a_2}$\qquad $(\textbf{1},\six,\textbf{1})_{\frac{1}{3}}$ &$0$ &$7$ &$0$ &$0$ &$0$ &$-\frac{1}{2}$ &$0$ &$0$\\ \hline
  $\ov \Ysymm_{a_2}$\qquad $(\textbf{1},\sixb,\textbf{1})_{-\frac{1}{3}}$ &$0$ &$-7$ &$0$ &$0$ &$0$ &$\frac{1}{2}$ &$0$ &$0$\\ \hline
  $\Yasymm_{a_2}$\qquad $(\textbf{1},\thrb,\textbf{1})_{\frac{1}{3}}$ &$0$ &$-1$ &$0$ &$0$ &$0$ &$-\frac{1}{2}$ &$0$ &$0$\\ \hline
  $\ov \Yasymm_{a_2}$\qquad $(\textbf{1},\thr,\textbf{1})_{-\frac{1}{3}}$ &$0$ &$1$ &$0$ &$0$ &$0$ &$\frac{1}{2}$ &$0$ &$0$\\ \hline
  $(\ov{a_1},a_2)$\qquad $(\thrb,\thr,\textbf{1})_{0}$ &$-3$ &$3$ &$0$ &$0$ &$-\frac{1}{2}$ &$\frac{1}{2}$ &$0$ &$0$\\ \hline
  $(a_1,\ov{a_2})$\qquad $(\thr,\thrb,\textbf{1})_{0}$ &$3$ &$-3$ &$0$ &$0$ &$\frac{1}{2}$ &$-\frac{1}{2}$ &$0$ &$0$\\ \hline
  $(a_1,a_2)$\qquad $(\thr,\thr,\textbf{1})_{\frac{1}{3}}$ &$3$ &$3$ &$0$ &$0$ &$-\frac{1}{2}$ &$-\frac{1}{2}$ &$0$ &$0$\\ \hline
  $(\ov{a_1},\ov{a_2})$\qquad $(\thrb,\thrb,\textbf{1})_{-\frac{1}{3}}$ &$-3$ &$-3$ &$0$ &$0$ &$\frac{1}{2}$ &$\frac{1}{2}$ &$0$ &$0$\\ \hline
  $(\ov{b},c)$\qquad $(\one,\textbf{1},\textbf{2})_{\frac{1}{2}}$ &$0$ &$0$ &$-1$ &$2$ &$0$ &$0$ &$-\frac{1}{2}$ &$-\frac{1}{3}$\\ \hline
  $(b,\ov{c})$\qquad $(\one,\textbf{1},\textbf{2})_{-\frac{1}{2}}$ &$0$ &$0$ &$1$ &$-2$ &$0$ &$0$ &$\frac{1}{2}$ &$\frac{1}{3}$\\ \hline
  $(b,c)$\qquad $(\one,\textbf{1},\textbf{2})_{\frac{1}{2}}$ &$0$ &$0$ &$1$ &$2$ &$0$ &$0$ &$-\frac{1}{2}$ &$-\frac{1}{3}$\\ \hline
  $(\ov{b},\ov{c})$\qquad $(\one,\textbf{1},\textbf{2})_{-\frac{1}{2}}$ &$0$ &$0$ &$-1$ &$-2$ &$0$ &$0$ &$\frac{1}{2}$ &$\frac{1}{3}$\\ \hline
  $\Ysymm_{b}$\qquad $(\one,\textbf{1},\textbf{3})_{0}$ &$0$ &$0$ &$6$ &$0$ &$0$ &$0$ &$0$ &$0$\\ \hline
  $\ov \Ysymm_{b}$\qquad $(\one,\textbf{1},\textbf{3})_{0}$ &$0$ &$0$ &$-6$ &$0$ &$0$ &$0$ &$0$ &$0$\\ \hline
  $\Yasymm_{b}$\qquad $(\one,\textbf{1},\textbf{1})_{0}$ &$0$ &$0$ &$-2$ &$0$ &$0$ &$0$ &$0$ &$0$\\ \hline
  $\ov \Yasymm_{b}$\qquad $(\one,\textbf{1},\textbf{1})_{0}$ &$0$ &$0$ &$2$ &$0$ &$0$ &$0$ &$0$ &$0$\\ \hline
  $\Ysymm_{c}$\qquad $(\one,\textbf{1},\textbf{1})_{1}$ &$0$ &$0$ &$0$ &$5$ &$0$ &$0$ &$0$ &$-\frac{4}{3}$\\ \hline
  $\ov \Ysymm_{c}$\qquad $(\one,\textbf{1},\textbf{1})_{-1}$ &$0$ &$0$ &$0$ &$-5$ &$0$ &$0$ &$0$ &$\frac{4}{3}$\\ \hline
\end{tabular}}
\caption{All possible fields for four-node axigluon quivers with $U(1)_Y = \frac{1}{6}U(1)_{a_1} + \frac{1}{6}U(1)_{a_2} + \frac{1}{6}U(1)_c$. The first column gives the representation of the field in the quiver spectrum and also under $\Ga\equiv SU(3)_L \times SU(3)_R \times SU(2)_L \times U(1)_Y$. The other columns give the T-charges and M-charges of that field. }
\label{table:quiver reps}

\end{table}

\clearpage

\bibliographystyle{JHEP}
\bibliography{axigluons}

\end{document}